\documentclass[twocolumn]{autart}    
\pdfoutput=1
\usepackage{graphicx}          
\usepackage{amssymb,amsmath,mathabx,amsfonts,mathrsfs,dsfont,cases}
\usepackage{tikz,adjustbox,booktabs,color}
\usepackage{hyperref}
\usepackage{lipsum}
\usepackage{natbib}
\begin{document}
\begin{frontmatter}
\runtitle{Beacon-referenced Cyclic Pursuit}  
\title{Collective Motion under Beacon-referenced Cyclic Pursuit} 
\author[KSG]{Kevin S. Galloway}\ead{kgallowa@usna.edu},    
\author[BD]{Biswadip Dey}\ead{biswadip@princeton.edu}
\address[KSG]{Electrical and Computer Engineering Department, United States Naval Academy, Annapolis, MD 21402, USA}
\address[BD]{Department of Mechanical and Aerospace Engineering, Princeton University, Princeton, NJ 08544, USA}
%
%
%
\begin{abstract}
Cyclic pursuit frameworks, which are built upon pursuit interactions between neighboring agents in a cycle graph, provide an efficient way to create useful global behaviors in a collective of autonomous robots. Previous work had considered cyclic pursuit with a constant bearing (CB) pursuit law, and demonstrated the existence of circling equilibria for the corresponding dynamics. In this work, we propose a beacon-referenced version of the CB pursuit law, wherein a stationary beacon provides an additional reference for the individual agents in a collective. When implemented in a cyclic framework, we show that the resulting dynamics admit relative equilibria corresponding to a circling orbit around the beacon, with the circling radius and the distribution of agents along the orbit determined by parameters of the proposed pursuit law. We also derive necessary conditions for stability of the circling equilibria, which provides a guide for parameter selection. Finally, by introducing a change of variables, we demonstrate the existence of a family of invariant manifolds related to spiraling motions around the beacon which preserve the ``pure shape'' of the collective, and study the reduced dynamics on a representative manifold.
\vspace{-1.5em}
\end{abstract}
%
%
%
\end{frontmatter}
%
%
%
\section{Introduction}
\label{sec:Intro}
%
A group of autonomous agents can accomplish certain missions more effectively and efficiently than individuals working alone, which may explain why such collective behaviors are often observed in nature \citep{Ballerini_TOPO, Scale, Fish_School_JTB, PigeonPaper} and increasingly implemented in robotic applications. Collective motion plays a pivotal role in modern robotics and engineering, especially in the area of search and rescue missions \citep{ Robotic_USAR_Cntrl}, surveillance \citep{Bethke09_GNC} and environmental monitoring \citep{Leonard_ZHANG_Monterey}. As one primary objective in this context has been to achieve control using information only about the local neighbors, dyadic pursuit interactions serve as an effective building block for collective motion \citep{Kim20071426, Marshall_TAC_04, 5160735, CollectiveMot_PrstEscp_Couzin, Sinha20071954}. Pursuit strategies are used in natural settings for capturing prey or pursuing a potential mate, and find application in robotic settings for rendezvous, missile defense, etc. Moreover, they provide an intuitive method for prescribing desired geometric relationships between autonomous agents, and can be executed by means of feedback-based pursuit laws such as the constant bearing (CB) pursuit law developed by \cite{Wei_Justh_PSK_09}. 

When a collective of agents implement CB pursuit in a cyclic setting (i.e. agent $i$ pursues agent $i+1$, with the last agent pursuing the first one), an earlier work by \cite{Galloway_PRS_13} has demonstrated existence of useful collective motions such as circling, spiraling, and rectilinear motion. For collectives of three agents, subsequent work established stability properties and revealed existence of trajectories that are periodic in shape and undergo precession in the physical space \citep{Galloway_PRS_16}. However, in this line of work, both the location of the circumcenter (with respect to an inertial frame) and the radius of the circular orbit were determined by initial conditions rather than control parameters.

In the current work, we employ a modified version of the CB control law, in which the pursuer is attentive to both a neighboring agent as well as to a stationary beacon. In some sense this beacon-referenced control law \citep[first introduced by][]{KSG_BD_ACC_2015} could be viewed as a ``conflicted" or ``distracted" CB pursuit law in which the agent attempts to simultaneously execute possibly conflicting pursuit strategies with respect to the neighboring agent and the beacon. Attention to the neighboring agent may represent a desire to maintain affiliation with a collective, while the beacon could represent an attractive food source (in biological settings) or a target of interest for an unmanned vehicle. In what follows, we consider an $n$-agent collective in which each agent $i$ employs this ``beacon-referenced'' CB control law with respect to agent $i+1$ and a common beacon. Although the control law itself does not specifically incorporate a desired station-keeping range from the beacon, we will demonstrate that when employed in a cyclic pursuit framework, circling equilibria exist which are centered on the beacon position and have a radius determined by the control parameters (rather than initial conditions). 

Beacon-referenced (or ``target-centric'') cyclic pursuit has also been addressed by \citet{Daingade2015AImplementation} and \citet{Mallik2015ConsensusApplications}. In these works, the authors employ a classical pursuit steering law with respect to a virtual point which lies along the line connecting the pursued neighboring agent to the beacon. While similar in spirit, our control law is fundamentally different in that it is based on constant bearing pursuit with respect to two targets rather than classical pursuit of a virtual point between the targets. In addition to relative equilibria, our work also investigates pure shape equilibria, i.e. motions that render the pure shape of the collective constant.

The main contribution of the current work is to develop conditions for existence and stability of circling equilibria as well as invariant manifolds corresponding to pure shape equilibria in beacon-referenced cyclic pursuit collectives. This work expands on our original analysis \cite[see][]{KSG_BD_ACC_2015,KSG_BD_ACC_2016} by deriving a stricter version of the necessary conditions for local stability of the circling equilibria (Section~\ref{sec:Local_Stability}) and by providing an analysis of the reduced dynamics on the invariant manifolds (Section~\ref{sec:reducedDynamics}). While our approach is motivated by the numerous robotic station-keeping applications which require autonomous agents to orbit a specified location while maintaining a fixed formation shape and scale (e.g. search and rescue, environmental sensing, etc.), we also note that this work may provide insights into the mechanisms underlying collective behavior observed in nature. For example, beacon-referenced cyclic pursuit may provide tools for modeling the ``explore-exploit" behavior observed in animal collectives (e.g. during honeybees' search for food sources [\citealt{Seeley_Behav_Ecol_91}]).

The paper proceeds as follows. In Section~\ref{sec:Model}, we state the dynamics governing a collective of autonomous agents interacting with a fixed beacon, and we derive appropriate shape variables for describing the relative states of the agents. In Section~\ref{sec:CL_Dyna} we present the beacon-referenced CB pursuit law, and develop the associated closed-loop shape dynamics which form the basis for the subsequent analysis. Section~\ref{sec:Rel_EQ_exist} details conditions for existence of circling equilibria, and in Section~\ref{sec:Local_Stability} we derive necessary conditions for local stability of the circling equilibria.  These necessary conditions may be used to guide parameter selection to avoid combinations that are known to result in instability. In Section~\ref{sec:Pure_Shape}, a change of variables is used to reveal the existence of a family of invariant submanifolds corresponding to spiral motions in the real space which maintain the ``pure shape" of the formation (i.e. the shape up to geometric similarity), and an analysis of the reduced dynamics on the manifold is presented in Section~\ref{sec:reducedDynamics}. 
%
%
%

%
%
\section{Problem Formulation}
\label{sec:Model}
%
As discussed by \cite{Galloway_PRS_13}, three key components are necessary to describe any decentralized algorithm for a group of agents (e.g. autonomous vehicles). Once the agents' dynamics have been described using appropriate \textit{generative models}, we specify the interaction structure by a directed \textit{attention graph}. Finally, we prescribe the \textit{feedback laws} governing the motion of individual agents. In what follows, we discuss each of these building blocks in the current context.
%
%

%
%
\subsection{Generative Model: Agents as Self-Steering Particles}
%
As we treat the agents as unit-mass self-steering particles on a plane \citep{Justh_PSK_SCL04}, natural Frenet frame equations \citep{Nat_Frenet_Bishop} provide a way to describe their motion. Then, by letting $\mathbf{r}_i$ and $\mathbf{x}_i$ denote the position and normalized velocity of the $i$-th agent, its dynamics can be expressed as 
\begin{equation}
\dot{\mathbf{r}}_i = \nu_i \mathbf{x}_i; \;
\dot{\mathbf{x}}_i = \nu_i u_i \mathbf{y}_i; \;
\dot{\mathbf{y}}_i = - \nu_i u_i \mathbf{x}_i,
\label{Explicit_MODEL}
\end{equation}
where $i \in \{1,\ldots,n\}$, $\mathbf{y}_i$ is the orthogonal rotation of $\mathbf{x}_i$ in the counter-clockwise direction, $\nu_i$ denotes speed, and $u_i$ is the natural curvature viewed as a steering control. We also introduce a stationary \textit{beacon} at position $\mathbf{r}_b$.
%
%

%
%
\subsection{Attention Graph}
%
Next, we define a directed graph $\mathcal{G} = (\mathcal{N}, \mathcal{A})$ with node set $\mathcal{N} = \{1,2,\ldots,n,b\}$. The associated arc set is defined as $\mathcal{A} = \{(i,i+1), (i,b) | i = 1,\ldots,n \}$\footnote{Addition in the index variables should be interpreted modulo $n$ throughout this paper.}. This weakly connected \textit{attention graph} \citep{Galloway_PRS_13} $\mathcal{G}$ captures the dyadic interactions in this problem.
%
%

%
%
\subsection{Reduction to Shape Space with Constraints}
\begin{figure}[t!]
\begin{center}
\includegraphics[width=0.45\textwidth]{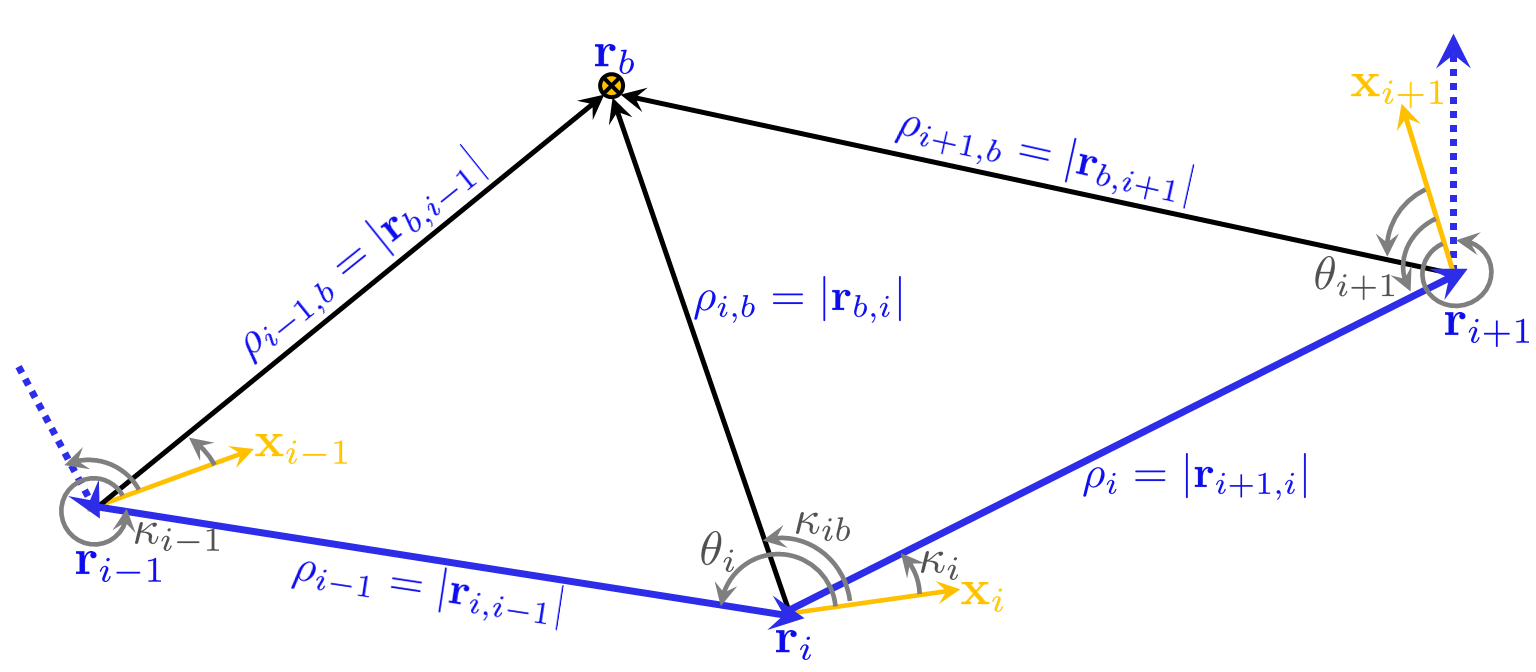}
\caption{Scalar shape variables associated with a beacon-referenced cyclic CB pursuit system with $n$ agents.}
\label{Scalar_Shapes}
\end{center}
\end{figure}

%
As our focus is towards studying the agents' motion relative to the beacon, we formulate a reduction to the shape space, i.e., the underlying space of relative positions and orientations. To do so,we first introduce a set of scalar shape variables (as shown in Figure~\ref{Scalar_Shapes}). By letting $R(\Omega)\in SO(2)$ denote a counter-clockwise rotation through an angle $\Omega$, we define the shape variables $\rho_i$, $\rho_{ib}$, $\kappa_i$, $\theta_i$, and $\kappa_{ib}$ as
\begin{align}
& \rho_i = |\mathbf{r}_{i+1,i}|,
&& R(\kappa_i)\mathbf{x}_i = \frac{\mathbf{r}_{i+1,i}}{|\mathbf{r}_{i+1,i}|}, 
&& R(\theta_i)\mathbf{x}_i = -\frac{\mathbf{r}_{i,i-1}}{|\mathbf{r}_{i,i-1}|}
\nonumber \\
& \rho_{ib} = |\mathbf{r}_{b,i}|,
&& R(\kappa_{ib})\mathbf{x}_i = \frac{\mathbf{r}_{b,i}}{|\mathbf{r}_{b,i}|},
&&
\label{Scalar_Shape_DEFN} 
\end{align}
where $i=1,\ldots,n$, and $\mathbf{r}_{i,j} = \mathbf{r}_{i} - \mathbf{r}_{j}$ represents the position of the $i$-th agent relative to the $j$-th agent. Although these variables ($5n$ in total) overparameterize the underlying space of relative position and orientation (of dimension $3n-1$), this effect can be taken into account by considering the constraints inherent to the system as follows.

From Figure~\ref{Scalar_Shapes}, one can observe that the normalized velocity of agent $i$ (i.e. $\mathbf{x}_i$) is related to $\mathbf{x}_{i+1}$ by a counter-clockwise rotation of $(\pi - \theta_{i+1} + \kappa_i)$. As successive application of such transformations over all agents yield an identity transformation, we have the following \textit{cycle closure constraint}:
\begin{equation}
R\big( \sum\limits_{i = 1}^{n} (\pi + \kappa_i - \theta_{i+1}) \big) = \mathds{I}_2.
\label{FINAL_Constraint_1}
\end{equation}
Furthermore, to maintain consistency, the vector sum of $\mathbf{r}_{i+1,i}$ and $\mathbf{r}_{b,i+1}$ should be the same as the baseline vector $\mathbf{r}_{b,i}$ between agent $i$ and the beacon. This observation leads to the following \textit{consistency condition}:
\begin{equation}
\rho_i \mathds{I}_2 = \rho_{ib}R(\kappa_{ib} - \kappa_i) + \rho_{i+1,b}R(\kappa_{i+1,b} - \theta_{i+1}),
\label{FINAL_Constraint_2}
\end{equation}
for $i = 1,\ldots,n$. This allows us to describe the shape space by the scalar shape variables $\{ \kappa_i, \kappa_{ib}, \theta_i, \rho_i, \rho_{ib} \}$, subject to (\ref{FINAL_Constraint_1}-\ref{FINAL_Constraint_2}). Additionally, we require $\rho_{i}$ to be positive for the shape variables \eqref{Scalar_Shape_DEFN} to be well-defined.
%
%
%

%
%
\section{Beacon-referenced CB pursuit law}
\label{sec:CL_Dyna}
%
Here we propose a beacon-referenced CB pursuit law, which we can express as a convex combination of two fundamental building blocks:
\begin{equation}
u_i = (1 - \lambda)u_{CB}^i + \lambda u_B^i, \qquad \lambda \in [0,1].
\label{u_i_top_level}
\end{equation}
Here $u_{CB}^i$ is given by the original CB pursuit law \citep{Wei_Justh_PSK_09} referenced to agent $i+1$, and $u_B^i$ represents the deviation from a desired bearing angle to the beacon, as will be made clear below. In particular, we define
\begin{align}
u_{CB}^i
&=
- \mu_i \left( R(\alpha_i)\mathbf{y}_i \cdot \frac{\mathbf{r}_{i,i+1}}{|\mathbf{r}_{i,i+1}|} \right)
\nonumber \\
& \qquad
- \frac{1}{\nu_i |\mathbf{r}_{i,i+1}|} \left( \frac{\mathbf{r}_{i,i+1}}{|\mathbf{r}_{i,i+1}|} \cdot R(\pi/2) \dot{\mathbf{r}}_{i,i+1} \right),
\label{u_CB_i}
\end{align}
where $\mu_i > 0$ is a control gain, and $\alpha_i$ is the desired offset between the $i$-th agent's heading and its bearing to agent $i+1$. We define the beacon tracking component as
\begin{equation}
u_B^i
=
- \mu_i^b \left( R(\alpha_{ib})\mathbf{y}_i \cdot \frac{\mathbf{r}_{i,b}}{|\mathbf{r}_{i,b}|} \right),
\label{u_Beacon_i}
\end{equation}
with $\mu_i^b > 0$ serving as the control gain. The angle $\alpha_{ib}$ represents the desired offset between the $i$-th agent's heading and its bearing to the beacon location. The parameter $\lambda$ maintains a balance between these two conflicting objectives, and $u_i$ simplifies to the original CB pursuit law whenever $\lambda = 0$. On the other extreme, the interaction between individual agents is completely lost for $\lambda = 1$. Therefore we restrict $\lambda$ to the open interval $(0,1)$ for the rest of our analysis. 

In terms of scalar shape variables, the feedback law \eqref{u_i_top_level} can be expressed as
\begin{align}
u_i
&=
\lambda \mu_i^b\sin(\kappa_{ib} - \alpha_{ib})
+
(1 - \lambda)\mu_i\sin(\kappa_i - \alpha_i) 
\nonumber \\
& \qquad 
+ \frac{1 - \lambda}{\rho_i}\left(\sin\kappa_i + \frac{\nu_{i+1}}{\nu_i}\sin\theta_{i+1}\right).
\label{u_i_shape}
\end{align}

\begin{rem}
As noted by \cite{Wei_Justh_PSK_09}, the last component of this feedback law \eqref{u_i_shape} is the angular speed at which the vector ${\bf r}_{i,i+1}$ is rotating around the agent $i$. Therefore it is plausible to compute $u_i$ without explicit range measurement, although it will require an appropriate sensing mechanism (mimicking the principle of compound eyes in visual insects).
\end{rem}

Before delving into further analysis, we introduce the following assumptions:\\
\indent(\textbf{A1}) The agents have constant and equal speed. Hence, without loss of generality, we can assume $\nu_i = 1$ for every $i = 1,\ldots,n$.\\
\indent(\textbf{A2}) The controller gains are common and equal for all agents, i.e. $\mu_i = \mu_i^b = \mu $, $i = 1,\ldots,n$.\\
\indent(\textbf{A3}) The bearing angles toward the beacon are equal for all agents, i.e. $\alpha_{ib} = \alpha_0 $, $i = 1,\ldots,n$.\\
\indent Such homogeneity can be enforced in a mobile vehicle context, and in a biological setting is not unreasonable to assume. We note that these assumptions still leave $n+3$ parameters to be chosen by the designer, allowing for a rich variety of system behaviors.
With these simplifying assumptions (A1)-(A3), the closed loop shape dynamics can be expressed as
\begin{align}
\dot{\rho}_i 
&= 
- \big( \cos\kappa_i + \cos\theta_{i+1} \big)
\nonumber \\
\dot{\kappa}_i 
&= 
- \mu \big[ (1 - \lambda) \sin(\kappa_i - \alpha_i) + \lambda \sin(\kappa_{ib} - \alpha_0) \big] 
\nonumber \\
& \qquad
+ \frac{\lambda}{\rho_i} \big[ \sin\kappa_i + \sin\theta_{i+1}\big]
\label{CL_dynamics_n_simplified} 
\\
\dot{\theta}_i 
&= 
\dot{\kappa}_i 
- \frac{1}{\rho_i}\big[ \sin\kappa_i + \sin\theta_{i+1}\big]
+ \frac{1}{\rho_{i-1}}\big[ \sin\kappa_{i-1} + \sin\theta_{i}\big]
\nonumber \\
\dot{\rho}_{ib} 
&= 
- \cos\kappa_{ib}
\nonumber \\
\dot{\kappa}_{ib} 
&= 
\dot{\kappa}_i 
- \frac{1}{\rho_i}\big[ \sin\kappa_i + \sin\theta_{i+1}\big]
+ \frac{1}{\rho_{ib}} \sin\kappa_{ib}
\nonumber
\end{align}
for $i = 1,\ldots,n$, subject to the cycle closure constraint \eqref{FINAL_Constraint_1} and consistency conditions \eqref{FINAL_Constraint_2}. It can be shown that the constraints \eqref{FINAL_Constraint_1}-\eqref{FINAL_Constraint_2} are preserved under the shape dynamics. The non-collocation constraint (i.e. $\rho_{i}>0$) is required for a well-defined control law but is not necessarily preserved by the shape dynamics, and therefore we restrict our analysis away from collision states.
%
%
%

%
%
\section{Relative equilibria: Circling motion}
\label{sec:Rel_EQ_exist}
\begin{figure}[t!]
\begin{center}
$\begin{array}{r}
\includegraphics[width=0.40\textwidth]{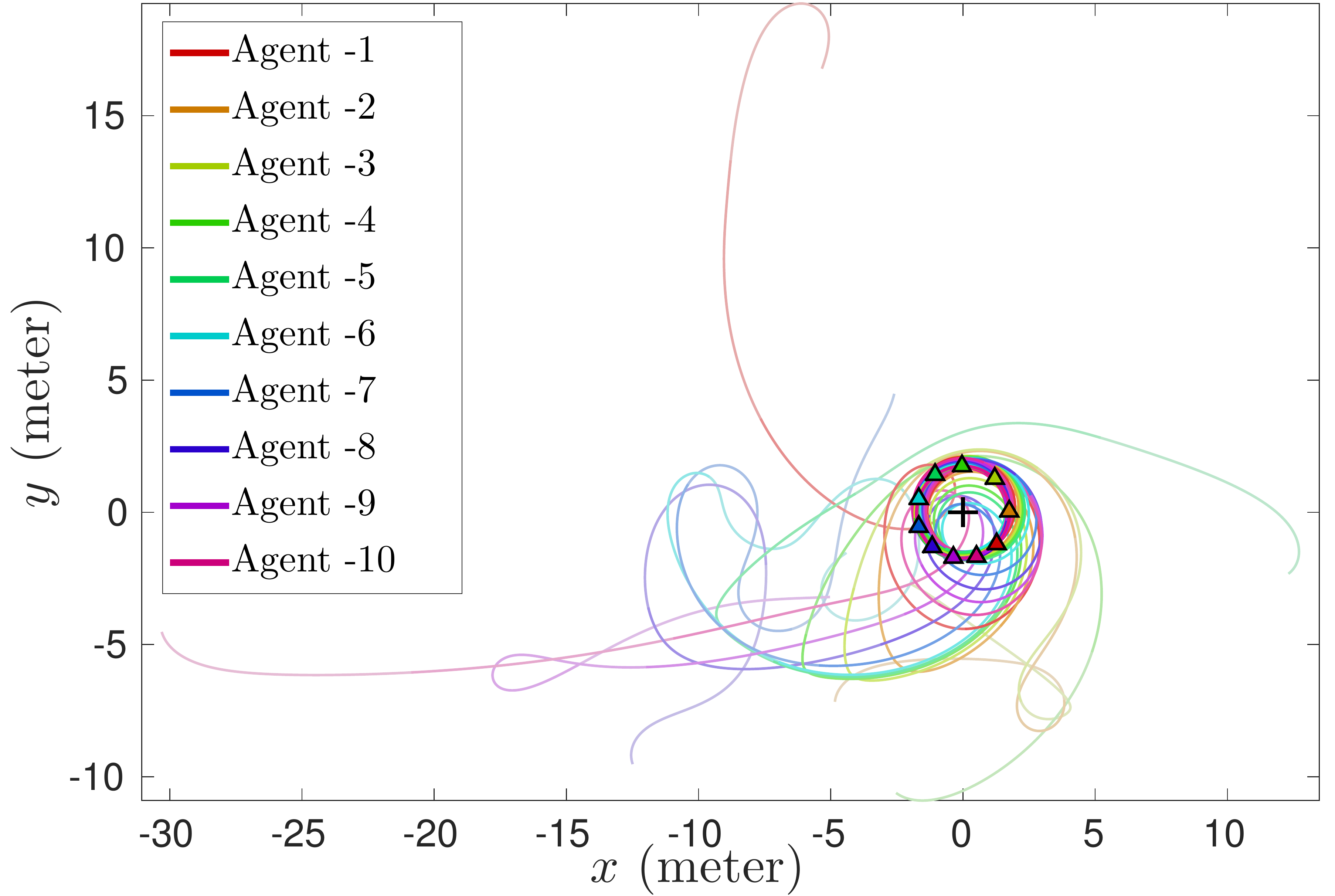}
\\
\includegraphics[width=0.39\textwidth]{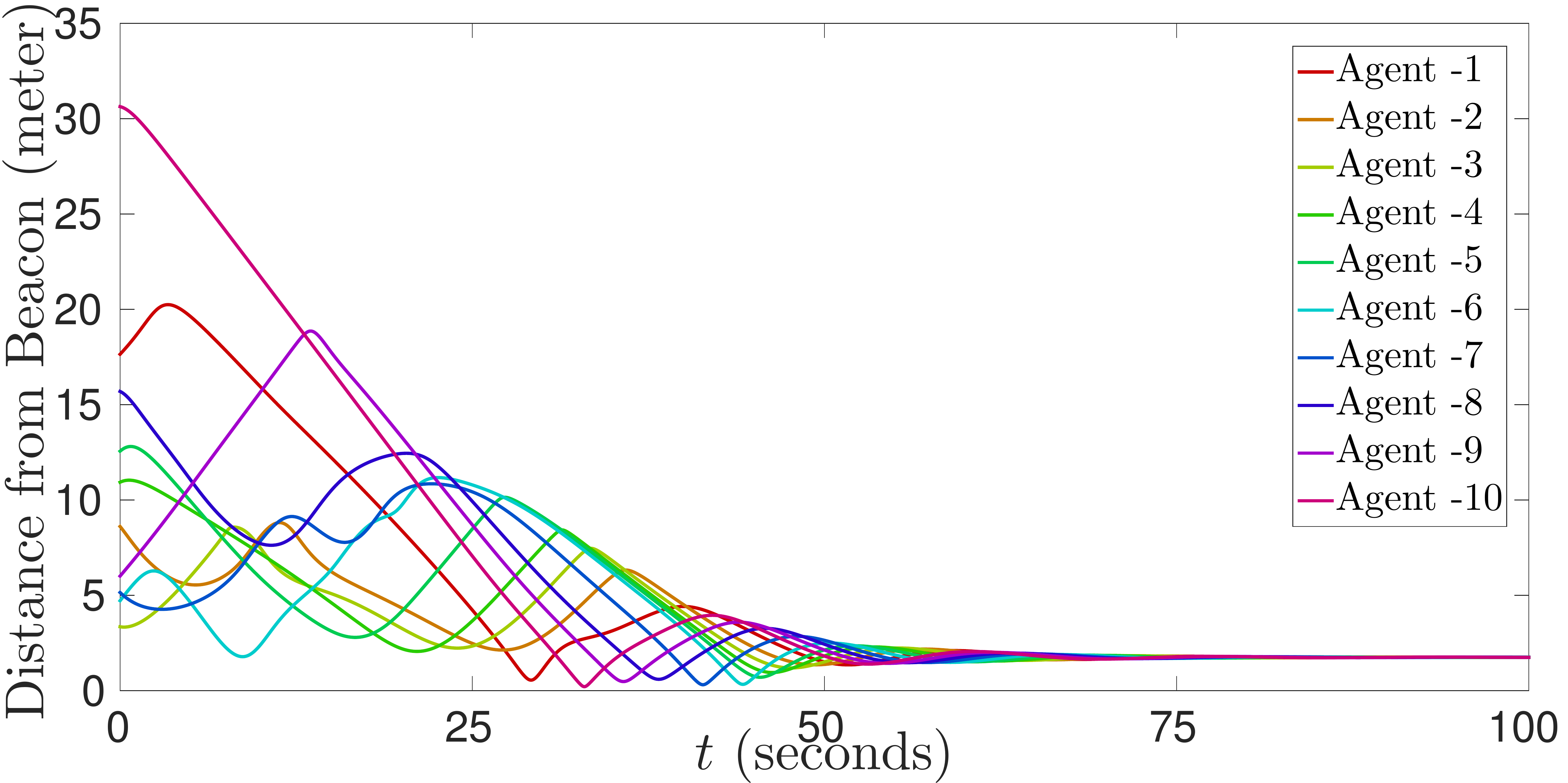}
\end{array}$
\caption{Matlab simulation results for a collective of 10 agents, with $\alpha_1 = \alpha_2 = \alpha_3 =\pi/6$, $\alpha_4 = \alpha_5 = \alpha_6 =\pi/7$, $\alpha_7 = \alpha_8 = \alpha_9 = \alpha_{10} =\pi/8$, $\alpha_0 = \pi/4$, $\mu = 1.0$, and $\lambda = 0.5$.}
\label{fig:circling}
\end{center}
\end{figure}

%
In this section we explore the equilibria of the closed loop shape dynamics \eqref{CL_dynamics_n_simplified}, which correspond to the relative equilibria of the original dynamics \eqref{Explicit_MODEL} with the beacon-referenced CB pursuit law \eqref{u_i_top_level}. We begin our analysis by setting the dynamics of $\rho_{ib}$ and $\rho_i$ to zero, and that leads to equilibria values of $\kappa_{ib}$ and $\theta_i$ given by
\begin{equation}
\kappa_{ib} = \pm \frac{\pi}{2},
\quad \textrm{and} \quad 
\theta_{i+1} = \pi \pm \kappa_i,
\quad 
i = 1,\ldots,n.
\label{eq_kappa_i_ib}
\end{equation}
Similarly, by setting the dynamics of $\theta_i$, $\kappa_i$, and $\kappa_{ib}$ to zero, we obtain
\begin{align}
\frac{1}{\rho_i}(\sin\kappa_i + \sin\theta_{i+1})
&=
\frac{1}{\rho_{i-1}}(\sin\kappa_{i-1} + \sin\theta_{i}),
\label{2nd_One}
\\
\frac{1}{\rho_i}(\sin\kappa_i + \sin\theta_{i+1})
&=
\frac{1}{\rho_{ib}} \sin\kappa_{ib},
\label{3rd_One}
\end{align}
for $i = 1,\ldots,n$. As the solution $\theta_{i+1} = \pi + \kappa_i$ leads to a contradiction in \eqref{3rd_One}, $\theta_i$ must satisfy
\begin{equation}
\theta_{i+1} = \pi - \kappa_i,
\quad 
i = 1,\ldots,n,
\label{theta_soln}
\end{equation}
at equilibria of the shape dynamics. Then, by introducing a new variable $\gamma_i$ defined as
\begin{equation}
\label{eqn:gammaDefn}
\gamma_i \triangleq \frac{1}{\rho_i}\big(\sin\kappa_i + \sin\theta_{i+1}\big) = \frac{2}{\rho_i} \sin\kappa_i,
\end{equation}
we obtain
\begin{equation}
\gamma_i = \gamma_{i-1}
\qquad 
i = 1,\ldots,n 
\label{eq_gamma_constant}
\end{equation}
from \eqref{2nd_One}. This condition, along with \eqref{3rd_One}, leads to
\begin{equation}
\gamma_i = \frac{\sin\kappa_{ib}}{\rho_{ib}} = \frac{\sin\kappa_{i-1,b}}{\rho_{i-1,b}}
\qquad 
i = 1,\ldots,n,
\label{relationship_agentbeacon_distances}
\end{equation}
and combined with \eqref{eq_kappa_i_ib}, we have 
\begin{equation}
\kappa_{ib}
=
\left\{
\begin{array}{ll}
 \pi/2 & \quad \forall i = 1,\ldots,n, \qquad \textrm{or} \\
-\pi/2 & \quad \forall i = 1,\ldots,n.
\end{array}
\right.
\label{kappa_ib_soln}
\end{equation}
Then it follows from \eqref{relationship_agentbeacon_distances} and \eqref{kappa_ib_soln}, that all agents will be equidistant from the beacon at any relative equilibrium. Hence, \textit{any relative equilibrium of the system must be a circling equilibrium}. Figure~\ref{fig:circling} depicts the results of a MATLAB simulation for a collective of 10 agents converging to such a circling equilibrium. (Control parameter specification is included in the figure caption.)

Next, by setting the dynamics of $\kappa_i$ to zero, we have
\begin{equation}
\mu \Big[(1 - \lambda)\sin(\kappa_i - \alpha_i) + \lambda \sin(\kappa_{ib} - \alpha_0) \Big]
= \lambda\gamma_i
\label{1st_One}
\end{equation}
for $i = 1,\ldots,n$. As $\kappa_{ib} = \pm\pi/2$, i.e. $\sin \kappa_{ib} = \pm 1$, at relative equilibria, \eqref{eqn:gammaDefn} and \eqref{1st_One} yield an equilibrium value for $\rho_i$, given by
\begin{equation}
\rho_i
=
\frac
{2 \lambda\sin\kappa_{ib} \sin\kappa_i}
{\mu \left( 1 - \lambda \right) \sin\kappa_{ib} \sin(\kappa_i - \alpha_i) + \mu \lambda \cos \alpha_0 }.
\label{eq_rho_i}
\end{equation}
Similarly, \eqref{3rd_One} yields an equilibrium value of $\rho_{ib}$ as
\begin{equation}
\rho_{ib}
=
\frac{\lambda}{\mu \left( 1 - \lambda \right) \sin\kappa_{ib} \sin(\kappa_i - \alpha_i) + \mu \lambda \cos \alpha_0}.
\label{eq_rho_ib_equidistant}
\end{equation}
Moreover, as both $\rho_{i}$ and $\rho_{ib}$ are required to be positive, 
\eqref{eq_rho_i} and \eqref{eq_rho_ib_equidistant} lead to the following \textit{necessary conditions} for existence of a circling equilibrium:
\begin{align}
&
\lambda \cos \alpha_0 + (1 - \lambda) \sin \kappa_{ib} \sin(\kappa_i - \alpha_i) > 0
\label{positive_1}
\\
\textrm{and,} \quad &
\sin \kappa_{ib} \sin\kappa_i > 0.
\label{positive_2}
\end{align}

The preceding discussion leads to the following theorem, which provides a complete characterization of the relative equilibria for the shape dynamics \eqref{CL_dynamics_n_simplified}. This was stated as Proposition~4.1 in our previous work \cite{KSG_BD_ACC_2015}, with a minor error corrected here in \eqref{eqn:equilibriumValuesFromTheorem}.
\begin{thm}
\label{prop:existenceProp}
Consider a beacon-referenced cyclic CB pursuit system with $n$ agents, and let its closed loop shape dynamics \eqref{CL_dynamics_n_simplified} be parametrized by $\mu$, $\lambda$, $\alpha_0$, and $\left\{\alpha_1, \ldots, \alpha_n \right\}$. Then, the following statements hold true: \\
\indent(a)
The only possible relative equilibria are circling equilibria.\\
\indent(b) Whenever $\sin(\sum \alpha_i) \neq 0$, a circling equilibrium exists \textit{if and only if} there exists $m \in \mathds{Z}$ and $\sigma = (\sigma_1,\sigma_2,\ldots,\sigma_n) \in \{-1,1\}^n$ such that \\
\indent \indent(i) 
the cardinality $M$ of the subset $\{\sigma_i | \sigma_i = 1, i = 1,\ldots,n\}$ satisfies
\begin{equation}
2M-n \neq 0,
\label{Prop_4_cond_on_M}
\end{equation}
and \\
\indent \indent(ii)
the following conditions hold true
\begin{equation}
\begin{aligned}
& \lambda \cos \alpha_0 + (1 - \lambda) \sin \alpha^* > 0,
\\
& \sin \big( \alpha^* + \sigma_i\alpha_i \big) > 0,
\end{aligned}
\label{Proposition_4}
\end{equation}
for $i = 1,\ldots,n$, where $\alpha^*$ is defined as
\begin{equation}
\alpha^* = \left( \frac{m+M-n}{2M-n} \right)\pi - \sum_{i=1}^{n} \left( \frac{\alpha_i}{2M-n} \right).
\label{alpha_star_soln}
\end{equation}
Moreover, at equilibrium, we have either $\kappa_{ib} = \pi/2, \; i=1,\ldots, n$ or $\kappa_{ib} = -\pi/2, \; i=1,\ldots, n$, and equilibrium values of $\kappa_i$, $\rho_{ib}$ and $\rho_i$ can be expressed as
\begin{equation}
\label{eqn:equilibriumValuesFromTheorem}
\begin{aligned}
\kappa_i &= \frac{\pi (1-\sigma_i)}{2} + (\sigma_i \alpha^* + \alpha_i) 
\\
\rho_{ib} &= \frac{\lambda}{\mu \lambda \cos \alpha_0 + \mu \left( 1 - \lambda \right) \sin\kappa_{ib} \sin\alpha^*}
\\
\rho_{i} &= 2\rho_{ib} \sin \kappa_{ib} \sin \kappa_i.
\end{aligned}
\end{equation}
\end{thm}
\begin{pf*}{Proof.}
Statement (a) of the theorem directly follows from \eqref{relationship_agentbeacon_distances} and \eqref{kappa_ib_soln}. Also, it follows from \eqref{theta_soln}, \eqref{eq_rho_i}, \eqref{eq_rho_ib_equidistant} and \eqref{positive_2}, that equilibrium values of $\rho_i$, $\rho_{ib}$, $\theta_i$ and $\kappa_{ib}$ can be expressed in terms of equilibrium values of $\kappa_i$. Now, in order to obtain a complete characterization of the relative equilibria, we focus on solving the equilibrium values of $\kappa_i$. Clearly, \eqref{1st_One} together with \eqref{eq_gamma_constant} leads to
\begin{equation}
\sin(\kappa_{i+1} - \alpha_{i+1})
=
\sin(\kappa_i - \alpha_i),
\label{angle_condition_n}
\end{equation}
at every relative equilibrium of the dynamics, and solving \eqref{angle_condition_n} we have
\begin{subnumcases}{\kappa_{i+1} - \alpha_{i+1} = }
\kappa_i - \alpha_i
\label{Solution_normal}
\\
\pi - (\kappa_i - \alpha_i)
\label{Solution_abnormal}
\end{subnumcases}
for $i = 1,\ldots,n$. Equilibrium values of $\kappa_i$ can therefore be obtained by solving \eqref{Solution_normal}-\eqref{Solution_abnormal} in conjunction with the shape variable constraints \eqref{FINAL_Constraint_1}-\eqref{FINAL_Constraint_2}.

Then, by letting $\alpha^*$ represent the bearing angle offset $(\kappa_1 - \alpha_1)$ at a relative equilibrium, \eqref{Solution_normal}-\eqref{Solution_abnormal} lead to either of the two possibilities for $(\kappa_2 - \alpha_2)$, namely $\alpha^*$ or $\pi - \alpha^*$. Furthermore, as this aspect of binary possibilities holds true for every agent, the possible solutions for \eqref{angle_condition_n} can be illustrated graphically via a binary tree (as shown in Figure~\ref{fig:tree_1}). Each branch in this binary tree represents a candidate solution for $\kappa_i$. In particular, the leftmost branch in this tree represents the relative equilibrium where $\kappa_i-\alpha_i = \kappa_{i+1}-\alpha_{i+1}$ for each $i=1,\ldots,n$. 

Taking \eqref{theta_soln} into consideration, we can express the cycle closure constraint \eqref{FINAL_Constraint_1} as
\begin{equation}
\sum_{i=1}^{n} \kappa_i = m\pi, 
\qquad m \in \mathds{Z},
\label{Closure_Bye_Product_1}
\end{equation}
at any relative equilibrium, i.e. equilibrium values of $\kappa_i$ must add up to an integral multiple of $\pi$. We now consider a representative branch of the binary tree (Figure~\ref{fig:tree_1}), along which (I) $\kappa_i - \alpha_i = \alpha^*$ for $M$ agents ($1 \leq M \leq n$), and (II) $\kappa_i - \alpha_i = \pi - \alpha^*$ for the remaining $n-M$ agents. Along this particular solution branch, \eqref{Closure_Bye_Product_1} can be expressed as
\begin{equation}
(2M-n)\alpha^*
= 
(m+M-n)\pi - \sum_{i=1}^n\alpha_i,
\label{cond_on_alpha_star}
\end{equation}
which in turn can be solved to obtain $\alpha^*$, as long as $2M - n \neq 0$. This leads to \eqref{alpha_star_soln}. Then, by introducing $\sigma \triangleq (\sigma_1,\ldots,\sigma_n) \in \{-1,1\}^n$ to denote whether an agent belongs to category I or II, we have $\kappa_i = (1-\sigma_i)\frac{\pi}{2} + \alpha_i + \sigma_i\alpha^*$ and $\sin\kappa_i = \sin( \alpha^*+ \sigma_i\alpha_i)$ for $i=1,\ldots,n$. Also, $M$ is the cardinality of the set $\{\sigma_i | \sigma_i = 1, i = 1,\ldots,n\}$.
%
%
\begin{figure}[t!]
\centering
\begin{tikzpicture}[
	level 1/.style={sibling distance=16mm},
	level 2/.style={sibling distance=14mm},
	level 3/.style={sibling distance=12mm},
	]
\node [circle,draw] (j) {$\alpha^*$}
  child {node [circle,draw] (k) {$\alpha^*$}
    child {node {$\vdots$}
      child {node [circle,draw] (o) {$\alpha^*$}}
      child {node [circle,draw] (p) {$\alpha_c^*$}}
    }
    child {node {$\vdots$}}
  }
  child {node [circle,draw] (l) {$\alpha_c^*$}
    child {node {$\vdots$}}
    child {node (c){$\vdots$}
      child {node [circle,draw] (o) {$\alpha_c^*$}}
      child {node [circle,draw] (p) {$\alpha^*$}
          child [grow=right] {node (q) {$\color{blue} \# n$} edge from parent[draw=none]
            child [grow=up] {node (r) {$\vdots$} edge from parent[draw=none]
              child [grow=up] {node (s) {$\color{blue} \# 2$} edge from parent[draw=none]
                child [grow=up] {node (t) {$\color{blue} \# 1$} edge from parent[draw=none]
              }
            }
          }
        }
      }
    }
  };
\end{tikzpicture}
\caption{Graphical representation of \textit{all possible} solution of \eqref{angle_condition_n}, where $\alpha_c^* = \pi - \alpha^*$.}
\label{fig:tree_1}
\end{figure}
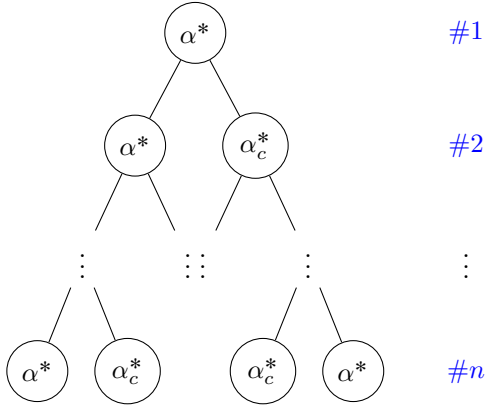
%

Whenever $\kappa_{ib} = \pi/2$, the positivity conditions \eqref{positive_1}-\eqref{positive_2} can readily be expressed as \eqref{Proposition_4}. Then, it remains to show that \eqref{Proposition_4} also encompasses the situation when $\kappa_{ib} = -\pi/2$, for which these conditions simplify to
\begin{equation}
\begin{aligned}
& 
\lambda \cos \alpha_0 - (1 - \lambda) \sin \alpha^* > 0
\\
&
\sin(\alpha^* + \sigma_i \alpha_i) < 0.
\end{aligned}
\label{CoNd__2}
\end{equation}
By introducing $\hat{m} \triangleq m+2M-n$, we can show that \eqref{Proposition_4}, with $\hat{m}$ substituted into \eqref{alpha_star_soln}, leads to \eqref{CoNd__2}. This establishes statement (b) of the theorem.
\qed
\end{pf*}
\begin{rem}
For an even number of agents, the possibility of having $2M-n= 0$ cannot be ruled out. In this case, from \eqref{cond_on_alpha_star} it follows that relative equilibria exist if and only if $\sum\alpha_i$ is an integer multiple of $\pi$, for which we have a continuum of relative equilibria of the shape dynamics.
\end{rem}
\begin{rem}
\label{remark:leftbranch}
Along the leftmost branch of the binary tree in Figure~\ref{fig:tree_1}, i.e. when \eqref{Solution_normal} holds true for each pair of agents, $\alpha^*$ can be expressed as
\begin{equation}
\alpha^* = m \left( \frac{\pi}{n} \right) - \sum_{i=1}^{n} \left( \frac{\alpha_i}{n} \right).
\label{alpha_soln_special_case}
\end{equation}
\end{rem}
\begin{rem}
From Theorem~\ref{prop:existenceProp} we can see that the $\alpha_i$ and $\alpha_{ib}$ parameters affect the system behavior in a much more complex (and less intuitive) way than is the case for cyclic CB pursuit without a beacon (as in \cite{Galloway_PRS_13}). However, the influence of these parameters can still be understood mathematically through Theorem~\ref{prop:existenceProp}, especially from the form of the equilibrium value for $\rho_{ib}$ in \eqref{eqn:equilibriumValuesFromTheorem} which shows how control parameters can be selected to specify a particular radius for the circling equilibrium.
\end{rem}
%
%
%

%
%
\section{Local Stability Analysis}
\label{sec:Local_Stability}
We proceed by introducing a simplifying assumption which will govern the analysis in the rest of this work:\\
\indent(\textbf{A4}) The bearing angles toward the neighbor are equal for all agents, i.e. $\alpha_{i} = \alpha$, $i = 1,\ldots,n$.

To investigate local stability of a relative equilibium, we first define $\xi_i \triangleq \{ \rho_i, \kappa_i, \theta_i, \rho_{ib}, \kappa_{ib} \}$, which allows us to express the shape dynamics \eqref{CL_dynamics_n_simplified} for agent-$i$ as $f(\xi_{i-1},\xi_i,\xi_{i+1})$. Then, by letting $\xi \triangleq \{\xi_1,\xi_2,\ldots,\xi_n\}$ denote the collective shape, we introduce $g_0(\xi) \triangleq \textstyle \sum_{i = 1}^{n} (\pi + \kappa_i - \theta_{i+1})$, $g_1^i(\xi) \triangleq \rho_i - \rho_{ib} \cos(\kappa_{ib} - \kappa_i) - \rho_{i+1,b} \cos(\kappa_{i+1,b} - \theta_{i+1})$, $g_2^i(\xi) \triangleq \rho_{ib} \sin(\kappa_{ib} - \kappa_i) + \rho_{i+1,b} \sin(\kappa_{i+1,b} - \theta_{i+1})$ for $i \in \{1,\ldots,n\}$. As these three functions express the shape variable constraints \eqref{FINAL_Constraint_1}-\eqref{FINAL_Constraint_2} as $g_0(\xi) = g_1^i(\xi) = g_2^i(\xi) =0$, the shape space $\mathcal{M} \subset \mathds{R}^{5n}$ can be defined as
\begin{displaymath}
\mathcal{M} = \{ \xi 
\; | \;
g_0(\xi) = g_1^i(\xi) = g_2^i(\xi) = 0, \forall i
\},
\end{displaymath}
and as discussed earlier $\mathcal{M}$ is invariant under the closed loop shape dynamics \eqref{CL_dynamics_n_simplified}. Therefore we focus our analysis on the dynamics restricted to this manifold $\mathcal{M}$.

Now we restrict our focus to a counter-clockwise\footnote{An analogous approach can be applied to the clockwise circling equilibrium.} circling equilibrium along the leftmost branch of the binary tree in Figure~\ref{fig:tree_1} (similar to Remark~\ref{remark:leftbranch}). Clearly, $\alpha^* = \frac{m\pi}{n} - \alpha$ and $\kappa_i = \frac{m\pi}{n}$, $\kappa_{ib} = \pi/2$, $i=1,\ldots,n$ for this equilibrium. Then, by letting $\bar{\xi} \triangleq \{ \bar{\xi}_1, \bar{\xi}_2, \ldots, \bar{\xi}_n \}$ represent this shape equilibrium, we introduce $\zeta_i \triangleq \xi_i - \bar{\xi}_i$ to denote a small perturbation. As the inter-agent interaction is cyclic in nature, the linearized dynamics around $\bar{\xi}_i$ can be expressed as
\begin{equation}
\dot{\zeta}_i = A_{_{0}}\zeta_{i} + A_{_{1}}\zeta_{i+1} + A_{_{-1}}\zeta_{i-1},
\end{equation}
where $A_{_{0}}, A_{_{1}}, A_{_{-1}} \in \mathds{R}^{5\times 5}$ are defined as
\begin{align*}
A_{_{0}}  &=
\left[\begin{array}{rrrrr}
0                & \sin\left(\frac{m\pi}{n}\right) & 0    & 0   & 0   \\
-\lambda q_1     & \lambda q_2 - q_3               & 0    & 0   & q_5 \\
(1-\lambda)q_1 & -(1-\lambda)q_2 - q_3         & -q_2 & 0   & q_5 \\
0                & 0                               & 0    & 0   & 1   \\
(1-\lambda)q_1 & -(1-\lambda)q_2 - q_3         & 0    & q_4 & q_5 
\end{array} \right]
\\
A_{_{1}}  &=
\left[\begin{array}{rrrrr}
\multicolumn{5}{c}{0_{2\times 5}} \\
\sin\left(\frac{m\pi}{n}\right) &  -\lambda q_2 & (1-\lambda)q_2 & 0 & (1-\lambda)q_2 \\
\multicolumn{5}{c}{0_{2\times 5}} 
\end{array} \right]^T
\\
A_{_{-1}} &=
\left[\begin{array}{rrrrr}
\multicolumn{5}{c}{0_{2\times 5}} \\
-q_1 & q_2 & 0 & 0 & 0 \\
\multicolumn{5}{c}{0_{2\times 5}}
\end{array} \right],
\end{align*}
\begin{displaymath}
\begin{aligned}
\textrm{and,} \quad q_1 &= \frac{\mu^2}{2} \left( \cos\alpha_0 + \big(\frac{1}{\lambda}-1\big) \sin\alpha^*\right)^2 \csc\left(\frac{m\pi}{n}\right)
\\
q_2 &= \frac{\mu}{2} \left( \cos\alpha_0 + \big(\frac{1}{\lambda}-1\big) \sin\alpha^*\right) \cot\left(\frac{m\pi}{n}\right)
\\
q_3 &= \mu(1-\lambda)\cos\alpha^*
\\
q_4 &= -2 q_1 \sin\left(\frac{m\pi}{n}\right)
\\
q_5 &= -\mu\lambda\sin\alpha_0.
\end{aligned}
\end{displaymath}

Then by representing the complete shape dynamics as $\dot{\xi} = \mathbf{F}(\xi)$, its linearization around an equilibrium $\bar{\xi}$ can be expressed as $\dot{\zeta} = \hat{A}\zeta$, where $\hat{A}$ is defined as
\begin{equation}
\hat{A} 
\triangleq 
\left.\frac{\partial\mathbf{F}}{\partial\xi}\right|_{\bar{\xi}}
=
\textrm{circ}\big( A_{_{0}},A_{_{1}},0_{_{5\times 5}}, 0_{_{5\times 5}}, \cdots, A_{_{-1}} \big),
\end{equation}
with \emph{circ} denoting a block circulant matrix. As discussed in \cite[Appendix~A]{KSG_BD_ACC_2016}, the shape variable constraints yield $(2n+1)$ imaginary axis eigenvalues, and therefore we can characterize the local stability of the equilibrium in terms of the remaining $(3n-1)$ eigenvalues. The following proposition (stated as Proposition~3.1 in \cite{KSG_BD_ACC_2016}) characterizes the eigenvalues of $\hat{A}$.
\begin{prop}
The eigenvalues of $\hat{A}$ are given by the union of the eigenvalues of the matrices 
\begin{equation}
D_{_k} = A_{_{0}} + \omega^{k} A_{_{1}} + \omega^{-k} A_{_{-1}}
\label{D_i_defn}
\end{equation}
for $k = 0,\ldots,n-1$, and $\omega = e^{2\pi j/n}$ denotes the $n$-th root of unity.
\label{Lemma_A_hat_SPECTRUM}
\end{prop}
\begin{pf*}{Proof.}
As described by \cite{Davis_Book}, we can write $\hat{A}$ as 
\begin{displaymath}
\hat{A} = (F_n \otimes \mathbb{I}_m)^* \textrm{diag}(D_{_{0}},D_{_{1}}, \cdots, D_{_{n-1}}) (F_n \otimes \mathbb{I}_m),
\end{displaymath}
where the $k$-th diagonal block $D_{_k}$ is given by \eqref{D_i_defn}. Moreover, $F_n$ is a $n\times n$ Fourier matrix given by $[F_n]_{kl} = \omega^{(k-1)(l-1)}$, and we can show that $(F_n \otimes \mathbb{I}_m)^*(F_n \otimes \mathbb{I}_m) = \mathbb{I}_{mn}$. Then it easily follows that the eigenvalues of $\hat{A}$ are the union of the eigenvalues of the $D_{_k}$ matrices.
\qed
\end{pf*}

To investigate the stability of $\hat{A}$, we proceed by calculating $P_k(x)$, the characteristic polynomial of $D_{_{k}}$, as
\begin{align}
& P_k(x) 
\nonumber \\
&= 
x^5 +
\mu \left[ b + \left(\frac{a}{2}\right) (1-\lambda) (1-\omega^k)\cot\left(\frac{m\pi}{n}\right) \right] x^4
\nonumber \\
& \; +
\left(\frac{\mu^2 a}{2}\right)  \left[ 2a + (1-\omega^k)d + \lambda a (1+\omega^k) \right] x^3
\nonumber \\
& \; +
\left(\frac{\mu^3 a^2}{2}\right) (1-\lambda)(1-\omega^k)\left(\cos\alpha^* + a\cot\left(\frac{m\pi}{n}\right)\right) x^2
\nonumber \\
& \;  + 
\mu^3 a^2 b x^2
+ \left(\frac{\mu^4 a^3}{2}\right)  \left[(1-\omega^k)d + \lambda a (1+\omega^k) \right] x
\nonumber \\
& \; +
\left(\frac{\mu^5 a^4}{2}\right) (1-\omega^k)(1-\lambda) \cos\alpha^*,
\label{P_k_Char_Poly}
\end{align}
where $k \in \{0,\ldots, n-1\}$, and $a$, $b$, $d$ are defined as
\begin{equation}
\begin{aligned}
a &= \cos\alpha_0 + (\frac{1}{\lambda}-1)\sin\alpha^*,
\\
b &= \lambda\sin\alpha_0 + (1-\lambda)\cos\alpha^*,
\\
\textrm{and,} \quad 
d &= a + (1-\lambda)\cos\alpha^*\cot(\frac{m\pi}{n}).
\end{aligned}
\label{abcd_DEFN}
\end{equation}
From \eqref{Proposition_4}, it is clear that $a$ should be positive for existence of the relative equilibrium $\bar{\xi}$. Furthermore, we can factorize each of these characteristic polynomials as
\begin{align}
P_k(x) &= \big( x^2 + \mu^2 a^2\big) \Big[
\big( x^3 + \mu \tilde{c}_k x^2 + \mu^2 a \tilde{d}_k x + \mu^3 a^2 \tilde{e}_k \big)
\nonumber \\
& \qquad \qquad 
-j \big( \mu \hat{c}_k x^2 - \mu^2 a \hat{d}_k x + \mu^3 a^2 \hat{e}_k \big)\Big],
\label{Char_Poly_k_Factor}
\end{align}
where $\tilde{c}_k$, $\hat{c}_k$, $\tilde{d}_k$, $\hat{d}_k$, $\tilde{e}_k$ and $\hat{e}_k$ are defined as
\begin{equation}
\begin{aligned}
\tilde{c}_k &= b + a(1-\lambda)\sin^2\left(\frac{k\pi}{n}\right)\cot\left(\frac{m\pi}{n}\right)
\\
\hat{c}_k &= a(1-\lambda)\sin\left(\frac{k\pi}{n}\right)\cos\left(\frac{k\pi}{n}\right)\cot\left(\frac{m\pi}{n}\right)
\\
\tilde{d}_k &= d\sin^2\left(\frac{k\pi}{n}\right) + \lambda a \cos^2\left(\frac{k\pi}{n}\right)
\\
\hat{d}_k &= (\lambda a - d)\sin\left(\frac{k\pi}{n}\right)\cos\left(\frac{k\pi}{n}\right)
\\
\tilde{e}_k &= (1 - \lambda) \cos\alpha^*\sin^2\left(\frac{k\pi}{n}\right)
\\
\hat{e}_k &= (1 - \lambda) \cos\alpha^*\sin\left(\frac{k\pi}{n}\right)\cos\left(\frac{k\pi}{n}\right).
\end{aligned}
\label{poly_CoEff_Defn}
\end{equation}

\begin{thm}
\label{THM:MAIN_Stability}
Consider the counter-clockwise circling equilibrium $\bar{\xi}$ of the beacon-referenced cyclic pursuit system with $n$-agents. The following conditions must hold true for stability of this equilibrium:
\begin{equation}
\begin{aligned}
& \tilde{c}_k >0
\\
& \tilde{c}_k\big( \tilde{c}_k\tilde{d}_k - a\tilde{e}_k \big) - \hat{d}_k\big( \tilde{c}_k\hat{c}_k + a\hat{d}_k \big) >0
\\
& \Gamma_k^2 \tilde{e}_k + \Gamma_k\Lambda_k \hat{d}_k - \Lambda_k^2 \tilde{c}_k >0
\end{aligned}
\label{NeCeSSaRy_CONDn_MAIN}
\end{equation}
for each $k=0,\ldots,n-1$, where $\Gamma_k$ and $\Lambda_k$ are defined as
\begin{displaymath}
\begin{aligned}
\Gamma_k &= \tilde{c}_k \big( \tilde{c}_k\tilde{d}_k - \hat{c}_k\hat{d}_k \big) - a \big( \hat{d}_k\hat{d}_k + \tilde{c}_k\tilde{e}_k \big)
\\
\Lambda_k &= \tilde{c}_k \big( \hat{c}_k\tilde{e}_k - \tilde{c}_k\hat{e}_k \big) + a \hat{d}_k \tilde{e}_k .
\end{aligned}
\end{displaymath}
\end{thm}
\begin{pf*}{Proof.}
As the spectrum of $\hat{A}$ is given by the union of eigenvalues of individual diagonal blocks $D_{_{k}} = A_{_{0}} + \omega^{k} A_{_{1}} + \omega^{-k} A_{_{-1}}$, $\hat{A}$ will not have any eigenvalue on the right half plane if and only if the eigenvalues of $D_{_{k}}$ do not have any positive real part for each $k=0,\ldots,n-1$. Hence, the conditions under which the relevant eigenvalues of $\hat{A}$ will be on the left half plane will lead to necessary conditions for stability of the equilibrium $\bar{\xi}$.

It follows from \eqref{Char_Poly_k_Factor} that $P_k(x)$ has a pair of pure imaginary roots at $x=\pm j \mu a$ for each $k=0,\ldots,n-1$. As these pure imaginary roots correspond to the coordinate constraints, we shift our focus to uncover the conditions under which each root of the second cubic factor will be on the left half plane (LHP). 

Next, following the general Routh-like algorithm for complex polynomials, developed by \cite{Agashe_CMPLX_routh}, and by leveraging the fact that the controller gain $\mu$ is positive, we can show that the roots of the cubic factor will have strictly negative real part \textit{if and only if} \eqref{NeCeSSaRy_CONDn_MAIN} holds true. Therefore, \eqref{NeCeSSaRy_CONDn_MAIN} must be true for each $k$, for the equilibrium $\bar{\xi}$ to be a stable one.
\qed
\end{pf*}
\begin{cor}
Consider a beacon-referenced cyclic pursuit system with $n$-agents. Then, the bearing angle parameters $\alpha$ and $\alpha_0$ must satisfy
\begin{equation}
\lambda\sin\alpha_0 + (1-\lambda)\cos\left(\frac{m\pi}{n} - \alpha\right) >0,
\end{equation}
for stability of a counter-clockwise circling equilibrium.
\end{cor}
\begin{pf*}{Proof.}
By setting $k=0$ in \eqref{poly_CoEff_Defn}, we have $\tilde{c}_0 = b$, $\tilde{d}_0 = \lambda a$ and $\hat{c}_0 = \hat{d}_0 = \tilde{e}_0 = \hat{e}_0 = 0$. Then it follows from Theorem~\ref{THM:MAIN_Stability} that $b$ (defined by \ref{abcd_DEFN}) must be positive for stability of the counter-clockwise circling equilibrium.
\qed
\end{pf*}
\begin{cor}
Consider a beacon-referenced cyclic pursuit system with $n$-agents. Then, whenever $n$ is even, the following conditions\footnote{These conditions have been stated as Proposition~3.2 in the earlier work of \cite{KSG_BD_ACC_2016}.} must hold true for stability of a counter-clockwise circling equilibrium:
\begin{equation}
\begin{aligned}
&
\cos\alpha^* >0
\\ & 
\lambda\sin\alpha_0 + (1-\lambda)\left(\cos\alpha^* + a \cot(\frac{m\pi}{n}) \right) >0
\\ &
bd + a(1-\lambda) \left[ d\cot\left(\frac{m\pi}{n}\right) - \cos\alpha^* \right] >0.
\end{aligned}
\end{equation}
\end{cor}
\begin{pf*}{Proof.}
By setting $k=\frac{n}{2}$, it directly follows from Theorem~\ref{THM:MAIN_Stability}.
\qed
\end{pf*}
\begin{rem}
By exploiting the fact that $a$, $b$ and $d$ do not depend on $\mu$ \eqref{abcd_DEFN}, it can be inferred from \eqref{P_k_Char_Poly} that $\mu$ just provides a scaling for the eigenvalues of diagonal blocks $D_{_k}$. Then, as $\mu$ is assumed to be positive, it readily follows that the stability of $\hat{A}$ is not influenced by $\mu$.
\end{rem}
%
%
%

%
%
\section{An invariant manifold: Pure shape equilibria}
\label{sec:Pure_Shape}
\begin{figure}[t!]
\begin{center}
\includegraphics[width=0.325\textwidth]{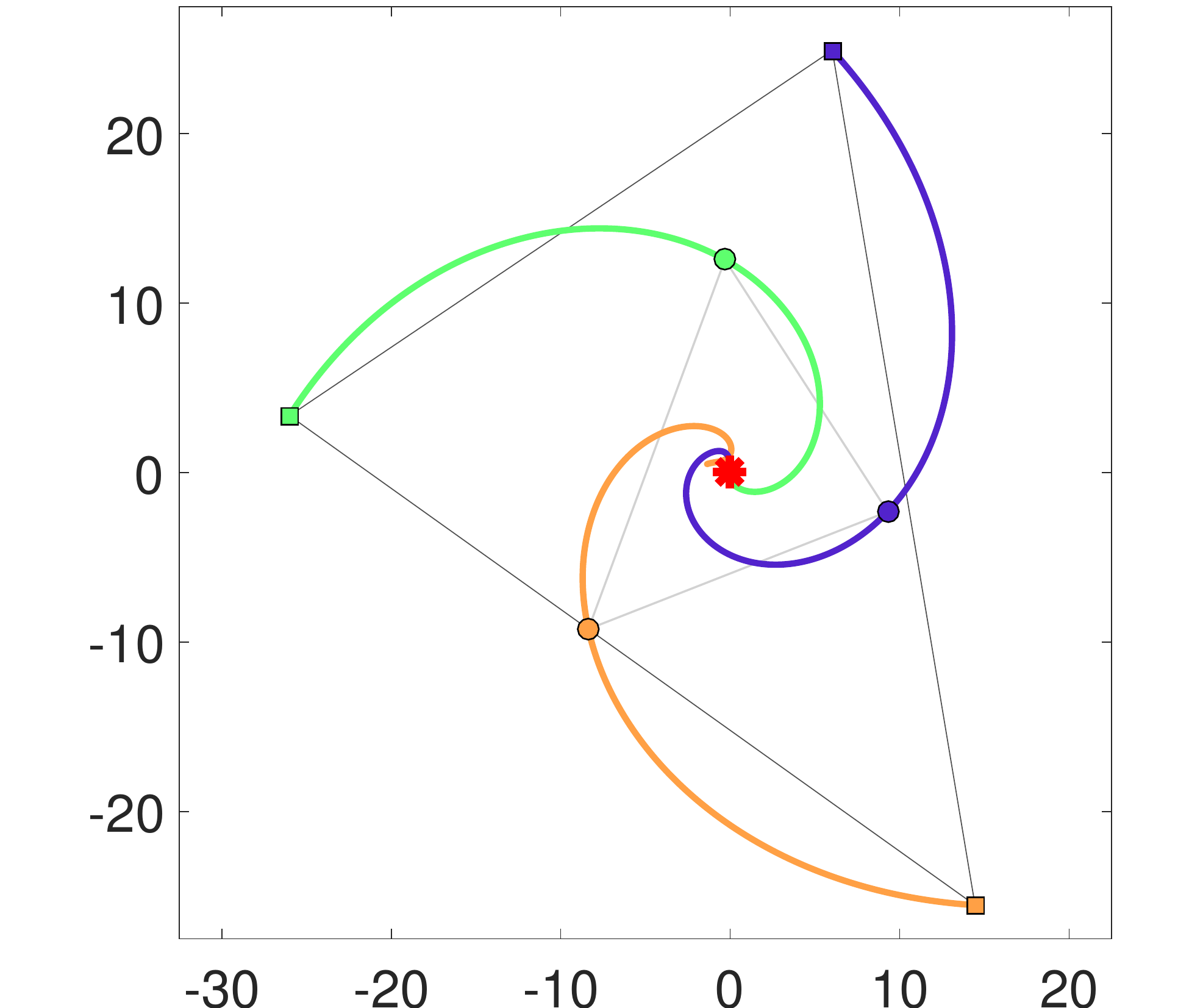}
\caption{This Matlab simulation shows spiraling out trajectories of the full space dynamics \eqref{Explicit_MODEL} for $n=3$, with $\alpha_0 = 11\pi/12$ and $\alpha_1 = \alpha_2 = \alpha_3 = 7\pi/12$.}
\label{fig:pureShapeTrajectories}
\end{center}
\end{figure}

%
In addition to circling equilibria, numerical simulations of the shape dynamics \eqref{CL_dynamics_n_simplified} indicate the existence of spiraling ``pure shape equilibria'' that maintain the shape of the collective up to geometric similarity (Figure~\ref{fig:pureShapeTrajectories}). In their previous work, \cite{Galloway_PRS_13} analyzed pure shape equilibria for cyclic pursuit systems (without a beacon) by means of rescaling the time variable. The time-scaling approach does not work in this beacon-referenced case, because the resulting pure shape dynamics are not self-contained (i.e. they depend on the size of the formation). However, we can use a change of variables to demonstrate the existence of a family of $(3n-3)$-dimensional invariant manifolds on which the pure shape remains constant. Note that while we continue to employ assumption (\textbf{A4}) for simplicity of presentation, preliminary results (to be presented in future work) indicate that conclusions analogous to Theorem \ref{prop:invariantManifold} hold for the case with heterogeneous $\alpha_i$ parameters.
%
%

%
%
\subsection{A change of variables}
%
We start with the shape variables developed in \eqref{Scalar_Shape_DEFN} and proceed by defining the following change of variables
\begin{equation}
\phi_{ib} \triangleq \kappa_{ib}-\kappa_i, \; 
\psi_i \triangleq \theta_i - \kappa_i, \; 
\tilde{\rho}_i \triangleq \frac{\rho_i}{\rho_1}, \; 
\tilde{\rho}_{ib} \triangleq \frac{\rho_{ib}}{\rho_1},
\end{equation}
for $i=1,\ldots,n$. Based on \eqref{FINAL_Constraint_1} and \eqref{FINAL_Constraint_2}, we have equivalent constraints on the new shape variables given by
\begin{align}
&
R\left( \sum\limits_{i = 1}^{n} (\pi - \psi_i) \right)
=
\mathds{I}_2
\label{FINAL_Constraint_NewVars_1}
\\
&
\tilde{\rho}_i \mathds{I}_2 = \tilde{\rho}_{ib}R(\phi_{ib}) + \tilde{\rho}_{i+1,b}R(\phi_{i+1,b} - \psi_{i+1}),
\label{FINAL_Constraint_NewVars_2}
\end{align}
for $i=1,\ldots,n$. Pure shape equilibria correspond to configurations wherein $\phi_{ib}, \psi_i, \tilde{\rho}_i$, and $\tilde{\rho}_{ib}$ remain constant for every $i=1,\ldots,n$, and all $\kappa_i$ vary at the same rate. Therefore we also define 
\begin{equation}
	\tilde{\kappa}_i \triangleq \kappa_i - \kappa_{i+1} , \; i=1,\ldots,n,
\end{equation}
subject to the constraint
\begin{equation}
\sum\limits_{i = 1}^{n} \tilde{\kappa}_i = 0.
\label{eqn:kappaTildeConstraint}
\end{equation}

It can be shown that our shape dynamics \eqref{CL_dynamics_n_simplified} can be parametrized in an alternative (but equivalent) form in terms of the variables $\kappa_1, \rho_1$, and $\left\{\tilde{\kappa}_i, \psi_i, \phi_{ib}, \tilde{\rho}_i, \tilde{\rho}_{ib}\right\}_{i=1}^n$. First, we note that in terms of these variables we have
\begin{equation}
	\kappa_i = \kappa_1 + \sum\limits_{j = i}^{n} \tilde{\kappa}_j,
\label{eqn:Kappa_Tilde_to_Kappa}
\end{equation}
and to simplify notation, we will denote
\begin{equation}
	\kappa_i^+ \triangleq \kappa_i + \kappa_{i+1} = 2\kappa_1 + \tilde{\kappa}_i + 2\sum\limits_{j = i+1}^{n} \tilde{\kappa}_j.
\label{eqn:kappaPlusInTermsOfKappaTilde}
\end{equation}
We also denote
\begin{equation}
\Phi_i \triangleq \kappa_{i}^+ + \psi_{i+1}, \quad \Psi_i \triangleq \tilde{\kappa}_{i} - \psi_{i+1},
\label{eqn:PhiPsiDefn}
\end{equation}
so that, by appropriate sum-to-product trigonometric identities, the shape dynamics \eqref{CL_dynamics_n_simplified} can be expressed in terms of the new variables as 
{\small 
\begin{align}
\label{CL_dynamics_n_shape3}
\dot{\kappa}_1 
&= 
- \mu \big[ (1 - \lambda) \sin(\kappa_1 - \alpha) + \lambda \sin(\phi_{1b} + \kappa_1 - \alpha_0) \big] 
\nonumber \\ 
& \qquad 
+ \frac{2\lambda \sin\left(\frac{\Phi_1}{2}\right) \cos\left(\frac{\Psi_1}{2}\right)}{\rho_1}, 
\nonumber \\
\dot{\rho}_1 
&=
-2 \cos \left(\frac{\Phi_1}{2}\right) \cos \left(\frac{\Psi_1}{2}\right),
\nonumber 
\\
\dot{\tilde{\kappa}}_i
&= -2\mu \Bigg[ (1 - \lambda) \sin \left(\frac{\tilde{\kappa}_i }{2}\right) \cos \left(\frac{\kappa_i^+ - 2\alpha }{2}\right) 
\\ 
& \qquad 
+\lambda \sin \left(\frac{\phi_{ib} - \phi_{i+1,b} + \tilde{\kappa}_i}{2}\right) 
\nonumber \\
& \quad \qquad 
\times 
\cos \left(\frac{\phi_{ib} + \phi_{i+1,b} +\kappa_i^+ - 2\alpha_0}{2}\right) \Bigg]
\nonumber
\\
& \qquad 
+ \frac{2\lambda}{\rho_1} \left[ \frac{\sin \left(\frac{\Phi_i}{2}\right) \cos \left(\frac{\Psi_i}{2}\right)}{\tilde{\rho}_{i}} - \frac{\sin \left(\frac{\Phi_{i+1}}{2}\right) \cos \left(\frac{\Psi_{i+1}}{2}\right)}{\tilde{\rho}_{i+1}} \right], 
\nonumber 
\\
\dot{\tilde{\rho}}_i 
&= 
\frac{2}{\rho_1} \left[ \tilde{\rho}_i \cos \left(\frac{\Phi_1}{2}\right) \cos \left(\frac{\Psi_1}{2}\right) -\cos \left(\frac{\Phi_i}{2}\right) \cos \left(\frac{\Psi_i}{2}\right) \right],
\nonumber \\
\dot{\psi}_i
&=
\frac{2}{\rho_1} \left[ \frac{ \sin \left(\frac{\Phi_{i-1}}{2}\right) \cos\left(\frac{\Psi_{i-1}}{2}\right)}{\tilde{\rho}_{i-1}} - \frac{ \sin \left(\frac{\Phi_i}{2}\right) \cos \left(\frac{\Psi_i}{2}\right)}{\tilde{\rho}_{i}} \right], 
\nonumber \\
\dot{\tilde{\rho}}_{ib} 
&= 
\frac{1}{\rho_1} \Bigg[ 2\tilde{\rho}_{ib} \cos \left(\frac{\Phi_1}{2}\right) \cos \left(\frac{\Psi_1}{2}\right) 
\nonumber \\
& \quad \qquad 
- \cos\left(\phi_{ib}+ \kappa_1 + \sum\limits_{j = i}^{n} \tilde{\kappa}_j \right) \Bigg],
\displaybreak[4]
\nonumber \\
\dot{\phi}_{ib} 
&= 
\frac{1}{\rho_1} \Bigg[ \frac{1}{\tilde{\rho}_{ib}}\sin\left(\phi_{ib} + \kappa_1 + \sum\limits_{j = i}^{n} \tilde{\kappa}_j \right) 
- \frac{2 \sin \left(\frac{\Phi_i}{2}\right) \cos \left(\frac{\Psi_i}{2}\right)}{\tilde{\rho}_{i}}\Bigg],
\nonumber
\end{align}
}
for $i = 1,2,\ldots,n$, subject to the constraints \eqref{FINAL_Constraint_NewVars_1}-\eqref{FINAL_Constraint_NewVars_2} and the positivity constraint for $\rho_1$, $\tilde{\rho}_i$, and $\tilde{\rho}_{ib}$.
%
%

%
%
\subsection{An invariant submanifold}
%
We now demonstrate the existence of a family of invariant submanifolds for which all dynamics are zero except for $\dot{\kappa}_1$ and $\dot{\rho}_1$. To maintain tractability of the mathematical analysis, we make the following simplifying assumption and seek to establish sufficient (but not necessary) conditions for existence.\\
\indent(\textbf{A5}) Assume $\cos\left(\frac{\Phi_i}{2}\right) \neq 0$,  $\cos\left(\frac{\Psi_i}{2}\right)\neq 0$, and $\sin (\frac{\Phi_i}{2}) \neq 0$ for all $i=1,\ldots,n$.

We then proceed by setting $\dot{\tilde{\rho}}_i = 0$ to obtain
\begin{equation}
\frac{\tilde{\rho}_i}{\tilde{\rho}_{i-1}} 
= 
\frac{\cos (\frac{\Phi_i}{2}) \cos (\frac{\Psi_i}{2})}{\cos (\frac{\Phi_{i-1}}{2}) \cos (\frac{\Psi_{i-1}}{2})}, \; i=1,\ldots,n.
\label{eqn:tilde_rho_equilibrium_condition1b}
\end{equation}
Setting $\dot{\psi}_i = 0$ yields
\begin{equation}
\frac{\tilde{\rho}_i}{\tilde{\rho}_{i-1}} = \frac{\sin (\frac{\Phi_i}{2}) \cos (\frac{\Psi_i}{2})}{\sin (\frac{\Phi_{i-1}}{2}) \cos (\frac{\Psi_{i-1}}{2})}, \; i=1,\ldots,n,
\label{eqn:tilde_rho_equilibrium_condition2}
\end{equation}
so that equating \eqref{eqn:tilde_rho_equilibrium_condition1b} with \eqref{eqn:tilde_rho_equilibrium_condition2} yields (for all $i$)
\begin{equation}
\sin (\frac{\Phi_i}{2}) \cos (\frac{\Phi_{i-1}}{2}) - \cos (\frac{\Phi_i}{2}) \sin (\frac{\Phi_{i-1}}{2}) = 0.
\label{eqn:tilde_rho_equilibrium_condition3}
\end{equation}
Since \eqref{eqn:tilde_rho_equilibrium_condition3} can be expressed as
\begin{equation}
	\sin\left(\frac{\Phi_{i} - \Phi_{i-1}}{2}\right) = 0, \; i=1,\ldots,n,
\end{equation}
our resulting requirement is
\begin{equation}
\Phi_1 = \Phi_2 = \cdots = \Phi_n = \Phi
\label{eqn:PhiDefn}
\end{equation}
for some angle $\Phi$ not dependent on the index $i$. Then by \eqref{eqn:PhiPsiDefn} we have
\begin{equation}
\psi_{i} = \Phi - \kappa_{i-1}^+ , \; i=1,\ldots,n, 
\label{eqn:psiEquilibrium1}
\end{equation}
and by substitution into \eqref{FINAL_Constraint_NewVars_1} with some manipulation, we have $R \left( n \left(\pi - \Phi + \sum\limits_{i = 1}^{n} \frac{\kappa_{i}^+}{n}\right) \right) =\mathds{I}_2$, i.e. 
\begin{equation}
\pi - \Phi + \sum\limits_{i = 1}^{n} \frac{\kappa_{i}^+}{n} = \frac{2k\pi}{n}, \; \textrm{for some } k \in \{0,1,\ldots,n\}.
\label{eqn:PhiEquation1}
\end{equation}
Combining \eqref{eqn:psiEquilibrium1} with \eqref{eqn:PhiEquation1}, we have
\begin{equation}
\psi_i = \left(\frac{n-2k}{n}\right)\pi - \kappa_{i-1}^+ + \sum\limits_{j = 1}^{n} \frac{\kappa_{j}^+}{n}, \; i=1,\ldots,n.
\label{eqn:psiEquilibrium2}
\end{equation}

Before proceeding, we also note that substituting \eqref{eqn:PhiDefn} into \eqref{eqn:tilde_rho_equilibrium_condition2} implies
\begin{equation}
\frac{\tilde{\rho}_i}{\tilde{\rho}_{i-1}} = \frac{\cos (\frac{\Psi_i}{2})}{\cos (\frac{\Psi_{i-1}}{2})}, \; i=1,\ldots,n.
\label{eqn:tilde_rho_equilibrium_condition4}
\end{equation}

Now we set $\dot{\phi}_{ib}$ equal to zero, to arrive at 
\begin{equation}
\frac{\tilde{\rho}_{ib}}{\tilde{\rho}_i} 
=
\frac{\sin\left(\phi_{ib} + \kappa_1 + \sum\limits_{j = i}^{n} \tilde{\kappa}_j \right)}{2\sin (\frac{\Phi}{2}) \cos (\frac{\Psi_i}{2})}, \; i=1,\ldots,n,
\label{eqn:phiDotZero}
\end{equation}
and set $\dot{\tilde{\rho}}_{ib}$ equal to zero to obtain
\begin{equation}
\tilde{\rho}_{ib} 
=
\frac{\cos\left(\phi_{ib} + \kappa_1 + \sum\limits_{j = i}^{n} \tilde{\kappa}_j \right)}{2\cos (\frac{\Phi}{2}) \cos (\frac{\Psi_1}{2})}, \; i=1,\ldots,n.
\label{eqn:rhoBDotZero}
\end{equation}
For $i=1$, $\tilde{\rho}_1 = 1$ and equating \eqref{eqn:phiDotZero} with \eqref{eqn:rhoBDotZero} yields
\begin{equation}
\frac{\sin\left(\phi_{1b} + \kappa_1 \right)}{2 \sin (\frac{\Phi}{2}) \cos (\frac{\Psi_1}{2})} 
= 
\frac{\cos\left(\phi_{1b} + \kappa_1 \right)}{2 \cos (\frac{\Phi}{2}) \cos (\frac{\Psi_1}{2})},
\end{equation}
(where we have used \eqref{eqn:kappaTildeConstraint}), which in turn yields 
\begin{align}
\label{eqn:phi1Equilibrium}
\sin\left(\phi_{1b} + \kappa_1 - \frac{\Phi}{2}\right) = 0. 
\end{align}

For any $i$, \eqref{eqn:phiDotZero} with \eqref{eqn:rhoBDotZero} implies
\begin{align}
\tilde{\rho}_i 
&=
\frac{\sin (\frac{\Phi}{2}) \cos (\frac{\Psi_i}{2}) \cos\left(\phi_{ib} + \kappa_1 + \sum\limits_{j = i}^{n} \tilde{\kappa}_j \right)}{\cos (\frac{\Phi}{2}) \cos (\frac{\Psi_1}{2}) \sin\left(\phi_{ib} + \kappa_1 + \sum\limits_{j = i}^{n} \tilde{\kappa}_j \right)},
\label{eqn:rhoTildeExpression}
\end{align}
and thus by dividing \eqref{eqn:rhoTildeExpression} by the corresponding expression for $\tilde{\rho}_{i-1}$ and equating with \eqref{eqn:tilde_rho_equilibrium_condition4}, we have
\begin{displaymath}
\frac{\cos\Big(\phi_{ib} + \kappa_1 + \sum\limits_{j = i}^{n} \tilde{\kappa}_j \Big)
\sin\Big(\phi_{i-1,b} + \kappa_1 + \sum\limits_{j = i-1}^{n} \tilde{\kappa}_j \Big)}
{\sin\Big(\phi_{ib} + \kappa_1 + \sum\limits_{j = i}^{n} \tilde{\kappa}_j \Big)
\cos\Big(\phi_{i-1,b} + \kappa_1 + \sum\limits_{j = i-1}^{n} \tilde{\kappa}_j \Big)} 
= 1.
\end{displaymath}
Thus
\begin{align}
& \sin\Big(\phi_{ib} + \kappa_1 + \sum\limits_{j = i}^{n} \tilde{\kappa}_j - \Big(\phi_{i-1,b} + \kappa_1 + \sum\limits_{j = i-1}^{n} \tilde{\kappa}_j \Big)\Big)
=0, \nonumber \\
& \text{i.e.,} \sin\left(\phi_{ib} - \phi_{i-1,b} -\tilde{\kappa}_{i-1} \right) =0, \; i=1,\ldots,n.
\label{eqn:equilibriumPhiKappaCondition}
\end{align}
Therefore the angle quantity in \eqref{eqn:equilibriumPhiKappaCondition} must be a multiple of $\pi$, and we proceed by considering the case\footnote{Note that we do incur a loss of generality with several steps in the process, but our aim is only to derive sufficient conditions for existence of an invariant manifold related to pure shape equilibria. Future work will consider all the alternative options to determine necessary conditions as well.} where it is an even multiple of $\pi$, i.e. 
\begin{equation}
\phi_{ib} - \phi_{i-1,b} -\tilde{\kappa}_{i-1} = 0, \; i=1,\ldots,n.
\label{eqn:PhiEquilibriumCondition10}
\end{equation}
Under this assumption, setting $\dot{\tilde{\kappa}}_i = 0$ in \eqref{CL_dynamics_n_shape3} yields
\begin{equation}
-2\mu (1 - \lambda) \sin \left(\frac{\tilde{\kappa}_i }{2}\right) \cos \left(\frac{\kappa_i^+ - 2\alpha}{2}\right) = 0,
\label{eqn:KappaEquilibriumCondition1}
\end{equation}
since \eqref{eqn:tilde_rho_equilibrium_condition4} guarantees that the last term of $\dot{\tilde{\kappa}}_i$ is zero. We proceed by considering the case in which the sine term in \eqref{eqn:KappaEquilibriumCondition1} is zero, i.e. 
\begin{equation}
\tilde{\kappa}_i = 0, \; i=1,\ldots,n.
\label{eqn:KappaEquilibriumCondition2}
\end{equation}
Note from \eqref{eqn:kappaPlusInTermsOfKappaTilde} that this implies
\begin{equation}
\kappa_i^+ = 2\kappa_1,
\label{eqn:kappaPlusEquilibrium}
\end{equation}
and thus by substitution into \eqref{eqn:psiEquilibrium2}, we have
\begin{equation}
\psi_i = \left(\frac{n-2k}{n}\right)\pi, \; i=1,\ldots,n.
\label{eqn:psiEquilibrium3}
\end{equation}

Therefore by \eqref{eqn:PhiPsiDefn}, \eqref{eqn:KappaEquilibriumCondition2}, \eqref{eqn:kappaPlusEquilibrium} and \eqref{eqn:psiEquilibrium3} we have
\begin{align}
\label{eqn:PhiFinal}
\Phi = 2\kappa_1  + \left(\frac{n-2k}{n}\right)\pi, \;  \;
\Psi_i = \left(\frac{2k-n}{n}\right)\pi,
\end{align}
for all $i$. Substitution into \eqref{eqn:tilde_rho_equilibrium_condition4} yields
\begin{equation}
\frac{\tilde{\rho}_i}{\tilde{\rho}_{i-1}} 
= 
\frac{\cos\left(\frac{2k-n}{2n}\pi\right)}{\cos\left(\frac{2k-n}{2n}\pi \right)}
= 
1,
\label{eqn:tilde_rho_equilibrium_condition5}
\end{equation}
for $i = 1,2,\ldots,n$, and since $\tilde{\rho}_1 \equiv 1$, we have  
\begin{equation}
\tilde{\rho}_i = 1, \; i=1,\ldots,n.
\label{eqn:tilde_rho_equilibrium_condition6}
\end{equation}

Lastly, we note that substituting \eqref{eqn:KappaEquilibriumCondition2} into \eqref{eqn:PhiEquilibriumCondition10} yields
\begin{equation}
\phi_{ib} = \phi_{i-1,b}, \; i=1,\ldots,n.
\label{eqn:phiEquation2}
\end{equation}
We can determine an expression for $\phi_{1b}$ by returning to \eqref{eqn:phi1Equilibrium} and considering the case for which the angle quantity is an even multiple of $\pi$, for which
\begin{equation}
\phi_{1b} 
= -\kappa_1 + \frac{\Phi}{2} 
=\left(\frac{n-2k}{2n}\right)\pi,
\end{equation}
and combining with \eqref{eqn:phiEquation2} yields 
\begin{equation}
\phi_{ib} = \left(\frac{n-2k}{2n}\right)\pi, \; i=1,\ldots,n.
\label{eqn:phiEquation5}
\end{equation}
Lastly, substituting \eqref{eqn:KappaEquilibriumCondition2},\eqref{eqn:PhiFinal}, and \eqref{eqn:phiEquation5} into \eqref{eqn:rhoBDotZero} yields
\begin{equation}
\tilde{\rho}_{ib} 
=
\frac{1}{2\sin\left(\frac{k\pi}{n} \right)}, \; i=1,\ldots,n,
\end{equation}
requiring $k \neq 0,n$ to ensure that $\tilde{\rho}_{ib}$ is well-defined. We summarize our results in the following theorem, which appeared as Proposition 4.1 in our earlier work \citep{KSG_BD_ACC_2016}.
\begin{thm}
\label{prop:invariantManifold}
For any $k \in \left\{1,2,\ldots,n-1\right\}$, the manifold $\mathcal{M}_k$ defined by
\begin{align}
\mathcal{M}_{k} &\triangleq  
\Biggl\{\kappa_1, \rho_1, \left\{\tilde{\kappa}_i, \psi_i, \phi_{ib}, \tilde{\rho}_i, \tilde{\rho}_{ib}\right\}_{i=1}^n \bigg| \; \tilde{\kappa}_i = 0, \tilde{\rho}_i = 1,  
\nonumber \\
& \quad \qquad 
\psi_i = \left(\frac{n-2k}{n}\right)\pi, \; \phi_{ib} = \left(\frac{n-2k}{2n}\right)\pi, 
\nonumber \\	
& \quad \qquad
\tilde{\rho}_{ib} = \frac{1}{2\sin\left(\frac{k\pi}{n} \right)}	\Biggr\},
\label{eqn:invariantManifoldDefn}
\end{align}
is nonempty and invariant under \eqref{CL_dynamics_n_shape3}, with 2-dimensional reduced dynamics on the manifold given by
\begin{align}
\dot{\kappa}_1 
&= 
- \mu \left[ (1 - \lambda) \sin(\kappa_1 - \alpha) + \lambda \cos\left(\kappa_1-\frac{k\pi}{n}  - \alpha_0\right) \right] 
\nonumber \\
& \qquad 
+ \frac{2\lambda}{\rho_1} \cos\Bigl(\kappa_1  -\frac{k\pi}{n}  \Bigr) \sin\left(\frac{k\pi}{n} \right) 
\nonumber \\
\dot{\rho}_1 
&=
- \cos\kappa_1 +\cos\left(\kappa_1 - \frac{2k\pi}{n} \right).
\label{eqn:dynamics_on_manifold}
\end{align}
\end{thm}
\begin{pf*}{Proof.}
The invariance of $\mathcal{M}_k$ follows from the calculations above, so it remains to establish that the constraints associated with $\mathcal{M}_k$ satisfy the consistency constraints \eqref{FINAL_Constraint_NewVars_1}-\eqref{FINAL_Constraint_NewVars_2}. First note that substituting the $\mathcal{M}_k$ constraints into the left-hand side of \eqref{FINAL_Constraint_NewVars_1} yields
\begin{align*}
R\left( \sum\limits_{i = 1}^{n} (\pi - \psi_i) \right) 
= R\left( \sum\limits_{i = 1}^{n} \Big(\pi - \left(\frac{n-2k}{n}\right)\pi\Big) \right) 
= \mathds{I}_2,
\end{align*}
and therefore \eqref{FINAL_Constraint_NewVars_1} is satisfied. Next, substituting the $\mathcal{M}_k$ constraints into the right-hand side of \eqref{FINAL_Constraint_NewVars_2} and simplifying with the property $R(\theta) + R(-\theta) = 2\cos(\theta)\mathds{I}_2$ results in $\mathds{I}_2$. Since $\tilde{\rho}_i = 1$, this results matches the left-hand side of \eqref{FINAL_Constraint_NewVars_2}, and therefore the constraint is satisfied. The form for the reduced dynamics can be obtained by substitution of the manifold constraints into \eqref{CL_dynamics_n_shape3}.
\qed
\end{pf*}
%
%
%
%

%
%
\section{Analysis of reduced dynamics on the pure shape manifold}
\label{sec:reducedDynamics}
%
%
Theorem~\ref{prop:invariantManifold} describes existence conditions for invariance of a manifold under the dynamics \eqref{CL_dynamics_n_shape3}, with the 2-dimensional reduced dynamics on the manifold given by \eqref{eqn:dynamics_on_manifold}. In this section we analyze \eqref{eqn:dynamics_on_manifold} to determine the behavior of trajectories on the invariant manifold.

First, by sum-to-product trigonometric identities we can reformulate our $\rho_1$ dynamics from \eqref{eqn:dynamics_on_manifold} as
\begin{align}
\label{eqn:rho_dynamics_on_manifold_reformulated}
\dot{\rho}_1 
&= 
2\sin\left(\kappa_1 - \frac{k\pi}{n}\right)\sin\left(\frac{k\pi}{n}\right).
\end{align}
Since $k \in \left\{1,2,\ldots,n-1\right\}$ it holds that $\sin(k\pi/n) > 0$, and therefore equilibria for \eqref{eqn:dynamics_on_manifold} (if they exist) must satisfy $\sin(\kappa_1 - k\pi/n) = 0$. Substituting $\kappa_1 = k\pi/n$ into the $\dot{\kappa}_1$ equation and setting equal to zero yields 
\begin{align}
\rho_1 
&= 
\frac{2\lambda\sin\left(\frac{k\pi}{n} \right)}{\mu \left[ (1 - \lambda) \sin\left(\frac{k\pi}{n}  - \alpha\right) + \lambda \cos\left(\alpha_0\right) \right]}.
\label{eqn:rho1_equilibrium}
\end{align}
Analogous calculations with $\kappa_1 = k\pi/n + \pi$ yield the same equilibrium value for $\rho_1$, with the requirement (in both cases) that the denominator of \eqref{eqn:rho1_equilibrium} must be positive in order to maintain $\rho_1 > 0$. 

We are led to consider the case where the denominator of \eqref{eqn:rho1_equilibrium} is not positive and therefore circling equilibria do not exist. 

\begin{thm}
\label{prop:invariantRegionReducedDynamics}
Consider an $n$-agent system evolving on the manifold $\mathcal{M}_k$. If the control parameters satisfy 
\begin{align}
 (1 - \lambda) \sin\left(\frac{k\pi}{n}  - \alpha\right) + \lambda \cos\left(\alpha_0\right) \leq 0,
\label{eqn:invarianceCondition}
\end{align}
then the region 
\begin{align}
\Delta = \left\{(\kappa_1,\rho_1) : \kappa_1 \in \left(\frac{k\pi}{n},\frac{k\pi}{n} + \pi \right), \; \rho_1>0\right\}
\end{align}
is positively invariant under the dynamics \eqref{eqn:dynamics_on_manifold}, i.e. trajectories which enter or start in the region will stay in the region for all future time.
\end{thm}
\begin{pf*}{Proof.}
First, we note from \eqref{eqn:rho_dynamics_on_manifold_reformulated} that $\dot{\rho}_1 > 0$ on $\Delta$, and therefore trajectories can not leave the region through the $\rho_1 = 0$ boundary. Along the boundary where $\kappa_1 = \frac{k\pi}{n}$, we have (from \ref{eqn:dynamics_on_manifold})
\begin{align}
\dot{\kappa}_1 
&= 
- \mu \left[ (1 - \lambda) \sin\left(\frac{k\pi}{n} -\alpha \right) + \lambda \cos\left(\alpha_0\right) \right]  
\nonumber \\
& \qquad 
+ \frac{2\lambda}{\rho_1} \sin\left(\frac{k\pi}{n} \right)
\nonumber \\
&>0,
\end{align}
where we have used \eqref{eqn:invarianceCondition} and the fact that $k \in \left\{1,2,\ldots,n-1\right\}$. A similar set of calculations demonstrates that $\dot{\kappa}_1 < 0$ along the boundary where $\kappa_1 = \frac{k\pi}{n} + \pi$, which establishes the proposition.
\qed
\end{pf*}

To further understand the reduced dynamics, we simplify the analysis by the following choice of parameters:\\
\indent(\textbf{A6}) Let $\lambda =1/2$, $\mu=2$.\\
Under this assumption, \eqref{eqn:dynamics_on_manifold} becomes
\begin{align}
\dot{\kappa}_1 
&= 
-\sin(\kappa_1 - \alpha) - \sin\left(\kappa_1-\left(\frac{2k-n}{2n}\right)\pi  - \alpha_0\right) 
\nonumber \\
& \qquad 
+ \frac{1}{\rho_1} \cos\Bigl(\kappa_1  -\frac{k\pi}{n}  \Bigr) \sin\left(\frac{k\pi}{n} \right)
\nonumber \\
&= 
-2\sin\left(\kappa_1 - \gamma_{k,n}\pi - \alpha_0^+\right)
\cos\left(\gamma_{k,n}\pi + \alpha_0^-\right)   
\nonumber \\
& \qquad 
+ \frac{1}{\rho_1} \cos\Bigl(\kappa_1  -\frac{k\pi}{n}  \Bigr) \sin\left(\frac{k\pi}{n} \right),
\label{eqn:kappa1_dynamics_on_manifold_reformulated}
\end{align}
where 
\begin{align}
\gamma_{k,n} \triangleq \frac{2k-n}{4n}, \quad \alpha_0^+ \triangleq \frac{\alpha_0 + \alpha}{2}, \quad \alpha_0^- \triangleq \frac{\alpha_0 - \alpha}{2},
\label{eqn:gammaAlphaDefns}
\end{align}
and the denominator of \eqref{eqn:rho1_equilibrium} can be expressed as $2\cos\left(\gamma_{k,n}\pi - \alpha_0^+\right) \cos\left(\gamma_{k,n}\pi + \alpha_0^-\right)$.

Thus circling equilibria exist on $\mathcal{M}_k$ if and only if $\cos\left(\gamma_{k,n}\pi - \alpha_0^+\right) \cos\left(\gamma_{k,n}\pi + \alpha_0^-\right) > 0$, with equilibrium values given by \eqref{eqn:rho1_equilibrium} and $\kappa_1 = k\pi/n$ or $\kappa_1 = k\pi/n + \pi$. By linearization one can show that the $\kappa_1 = k\pi/n$ equilibrium is stable (and the $\kappa_1 = k\pi/n + \pi$ equilibrium is unstable) if $\sin\left(\gamma_{k,n}\pi - \alpha_0^+\right) \cos\left(\gamma_{k,n}\pi + \alpha_0^-\right) < 0$, with the opposite statement holding when this term is positive. This can be related to the conditions in Theorem~\ref{prop:existenceProp} for a particular value of $k$ by letting $\sigma_i = 1$ and $m=k$ or $m=k+n$ in part (b) of Theorem~\ref{prop:existenceProp}. 

\begin{figure}
\begin{center}
  \includegraphics[width=0.4\textwidth]{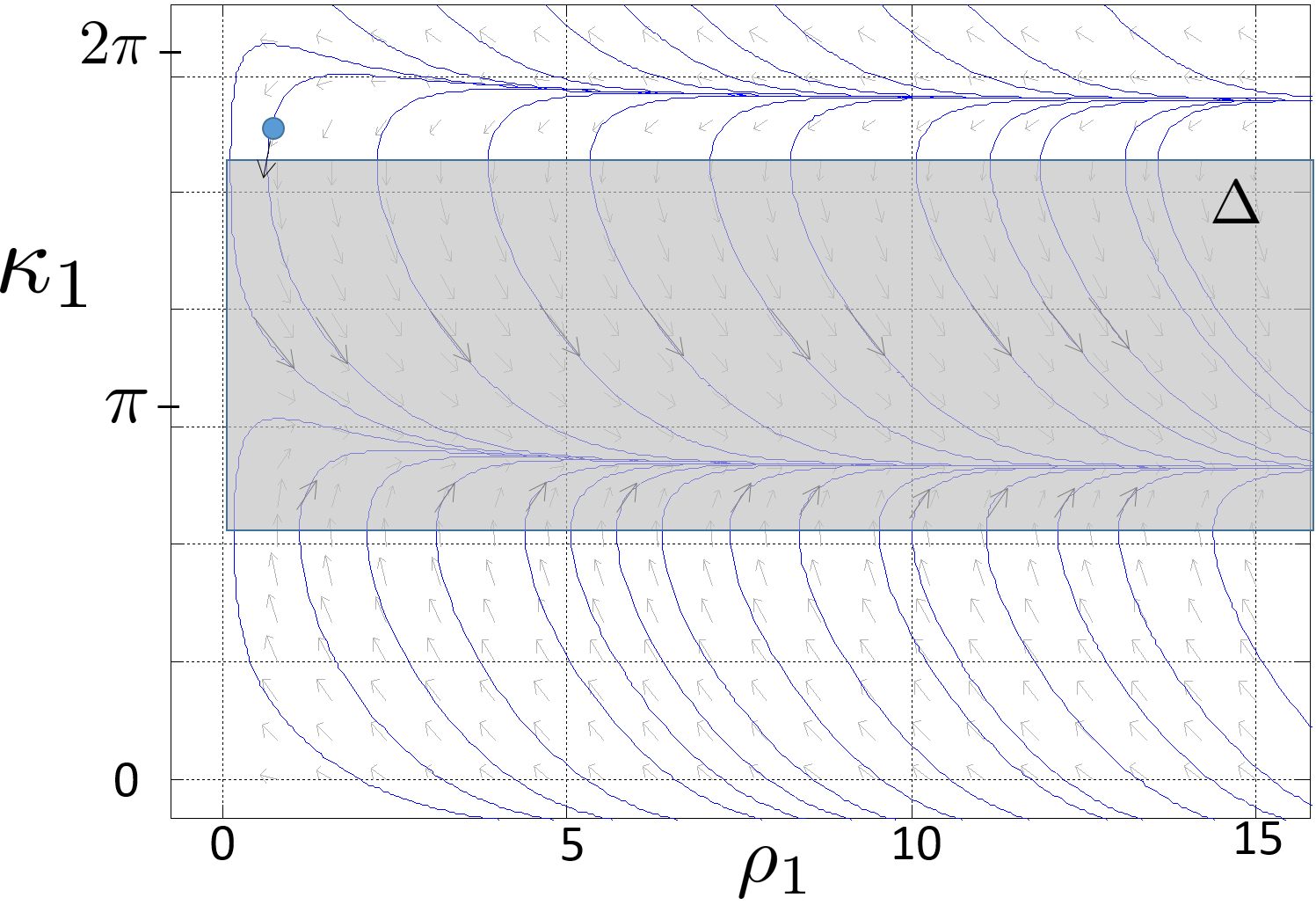}
  \caption{The $(\kappa_1, \rho_1)$ phase portrait for the dynamics \eqref{eqn:dynamics_on_manifold} with $n=3$, $k=2$, $\mu=2$, $\lambda=1/2$, $\alpha_0 = 11\pi/12$, and $\alpha = 7\pi/12$. The shaded area depicts the positively invariant region $\Delta$, with trajectories approaching the asymptote $\kappa_1 = \gamma_{2,3}\pi + \alpha_0^+ = 5\pi/6$. The trajectory starting at the solid circle corresponds to the spiraling motion in Figure~\ref{fig:pureShapeTrajectories}.} 
  \label{fig:phasePortrait}
\end{center}
\end{figure}

If $\cos\left(\gamma_{k,n}\pi - \alpha_0^+\right) \cos\left(\gamma_{k,n}\pi + \alpha_0^-\right) \leq 0$, then Theorem~\ref{prop:invariantRegionReducedDynamics} applies. In this case, the behavior of trajectories in the region $\Delta$ can be better analyzed by introducing a quantity $V_{k,n} \triangleq -\cos\left(\kappa_1 - \gamma_{k,n}\pi - \alpha_0^+ \right)$. Then differentiating $V_{k,n}$ along trajectories of \eqref{eqn:rho_dynamics_on_manifold_reformulated}-\eqref{eqn:kappa1_dynamics_on_manifold_reformulated}, we have
\begin{align*}
&\dot{V}_{k,n} 
= \dot{\kappa}_1 \sin\left(\kappa_1 - \gamma_{k,n}\pi - \alpha_0^+ \right)
\nonumber \\
& \;=
-2\sin^{2}\left(\kappa_1 - \gamma_{k,n}\pi - \alpha_0^+\right)
\cos\left(\gamma_{k,n}\pi + \alpha_0^-\right)   
\nonumber \\
& \quad 
+ \frac{1}{\rho_1} \sin\left(\kappa_1 - \gamma_{k,n}\pi - \alpha_0^+ \right)\cos\Bigl(\kappa_1  -\frac{k\pi}{n}  \Bigr) \sin\left(\frac{k\pi}{n} \right). 
\label{eqn:reducedDynamicsAsymptoteV}
\end{align*}
If $\cos\left(\gamma_{k,n}\pi + \alpha_0^-\right) >0$, then the first term in $\dot{V}_{k,n}$ will be strictly negative and will dominate for large values of $\rho_1$. We recall that $\dot{\rho}_1 > 0$ on $\Delta$, and eventually $\dot{V}_{k,n}$ will be strictly negative and decreasing towards $V_{k,n} = -1$, i.e. $\kappa_1$ will asymptotically approach the value $\gamma_{k,n}\pi + \alpha_0^+$. If $\cos\left(\gamma_{k,n}\pi + \alpha_0^-\right) <0$, then an analogous argument with $\hat{V}_{k,n} \triangleq \cos\left(\kappa_1 - \gamma_{k,n}\pi - \alpha_0^+ \right)$ demonstrates that $\kappa_1$ will approach the value $\gamma_{k,n}\pi + \alpha_0^+ + \pi$. These behaviors are depicted in the phase portrait representation in Figure~\ref{fig:phasePortrait}, which corresponds to the spiraling behaviors in the physical space depicted in Figure~\ref{fig:pureShapeTrajectories}. 

While the analysis presented here does not provide an exhaustive characterization of possible system trajectories on the invariant manifold $\mathcal{M}_k$, it does enable the use of phase portraits for further studies. As an example, phase portrait analysis of Figure~\ref{fig:phasePortrait} reveals the existence of trajectories that first spiral in before spiraling out. For other choices of parameters, phase portraits also suggest trajectories that are periodic in the phase variables.
\begin{rem}
Theorem~\ref{prop:invariantManifold} establishes the invariance of the manifold $\mathcal{M}_k$ but does not address the conditions under which the manifold is attractive, and the results of Theorem~\ref{prop:invariantRegionReducedDynamics} must be interpreted with this in mind. Numerical simulations of the full system dynamics (\eqref{Explicit_MODEL} with \eqref{u_i_top_level}) reveal the existence of other system trajectories (such as periodic trajectories with precession) which do not evolve on $\mathcal{M}_k$, suggesting that $\mathcal{M}_k$ is not attractive for all parameter choices.  An analysis of the attractivity of the subject manifold will be carried out in future work.   
\end{rem}
%
%
%

%
%
\section{Conclusion}
%
%
In this paper, we have proposed a modification of the constant bearing pursuit law which references a fixed beacon as well as a neighboring agent, and have demonstrated that implementation of such control law in a cycle graph (with ``spokes") yields an interesting set of closed-loop dynamics. Analysis of these dynamics revealed that the associated relative equilibria corresponds to circling of the agents around the beacon. We have also shown that the circling radius is determined by parameters of the control law, not by initial conditions (as was the case in earlier works on cyclic pursuit with constant bearing feedback law \cite{Galloway_PRS_13}). This independence from initial conditions has made this modified framework better suited for station-keeping applications. Then, after deriving necessary conditions for stability of the relative (circling) equilibria, we have characterized the invariant manifolds corresponding to spiral motions which preserve scale-invariant pure shape of the collective, and analyzed the motion on this invariant manifold. An experimental implementation of this framework was described in an earlier work by \cite{KSG_BD_ACC_2015}, and interested readers may refer to the implementation videos available at \url{http://ter.ps/beaconcb}.

Future work will analyze the attractivity of pure shape manifolds, and investigate the existence of more complex invariant submanifolds within the shape space suggested by numerical simulations. We also intend to expand our framework to 3-d settings, as well as consider multiple beacons or slowly moving beacons.
%
%
%
%
%

%
%
%
%
%
\end{document}